\documentclass[twocolumn]{aastex63} 
\usepackage{amsmath} 
\usepackage{hyperref} 
\usepackage{multirow} 
\usepackage{natbib}

\renewcommand{\vec}[1]{\mbox{\boldmath$#1$}}

\newcommand{\planck}{\textit{Planck}} 
\newcommand{\chandra}{\textit{Chandra}}

\newcommand{\yx}{\ensuremath{Y_\mathrm{X}}}

\newcommand{\asz}{\ensuremath{{A_\mathrm{SZ}}}} 
\newcommand{\bsz}{\ensuremath{{B_\mathrm{SZ}}}}
\newcommand{\csz}{\ensuremath{{C_\mathrm{SZ} \ }}} 
\newcommand{\sigmalnzeta}{\ensuremath{{\sigma_{\ln \zeta}}}} 
\newcommand{\ayx}{\ensuremath{{A_{\yx}}}}
\newcommand{\byx}{\ensuremath{{B_{\yx}}}} 
\newcommand{\cyx}{\ensuremath{{C_{\yx}}}} 
\newcommand{\sigmalnyx}{\ensuremath{{\sigma_{\ln \yx}}}}

\newcommand{\comment}[1]{} 
\newcommand{\Om}{\ensuremath{\Omega_\mathrm{m}}}

\newcommand{\sig}{\ensuremath{\sigma_8}} 
\newcommand{\seight}{\ensuremath{S_8}} 

\newcommand{\LCDM}{\ensuremath{\Lambda\mathrm{CDM}}} 
\newcommand{\lcdm}{\ensuremath{\Lambda\mathrm{CDM}}}
 
\newcommand{\wCDM}{\ensuremath{w\mathrm{CDM}}}

\newcommand{\sumMnu}{\ensuremath{\sum m_\nu}} 
\newcommand{\sqdeg}{\ensuremath{{\rm deg}^2}} 
\mathchardef\mhyphen="2D
\newcommand{\ukarcmin}{\ensuremath{\mu{\rm K\mhyphen arcmin}}}
\newcommand{\sigcmb }{\ensuremath{ 0.8118 \pm 0.0072}}
\newcommand{\seightcmb }{\ensuremath{ 0.834 \pm 0.016}}
\newcommand{\omcmb }{\ensuremath{ 0.3165 \pm 0.0084}}
\newcommand{\sigcurrent }{\ensuremath{ 0.737 \pm 0.033}}
\newcommand{\seightcurrent }{\ensuremath{ 0.794 \pm 0.049}}
\newcommand{\omcurrent }{\ensuremath{ 0.352 \pm 0.047}}
\newcommand{\aszcurrent }{\ensuremath{ 5.3 \pm 1.1}}
\newcommand{\bszcurrent }{\ensuremath{ 1.668 \pm 0.068}}
\newcommand{\cszcurrent }{\ensuremath{ 1.09 \pm 0.30}}
\newcommand{\dszcurrent }{\ensuremath{ 0.168 \pm 0.076}}
\newcommand{\sigcurrentwcmb }{\ensuremath{ 0.8081 \pm 0.0079}}
\newcommand{\seightcurrentwcmb }{\ensuremath{ 0.831 \pm 0.020}}
\newcommand{\omcurrentwcmb }{\ensuremath{ 0.316 \pm 0.011}}
\newcommand{\seightreductioncurrent }{\ensuremath{ 1.4}}
\newcommand{\wcdmsigcmb }{\ensuremath{ 0.985 \pm 0.077}}
\newcommand{\wcdmseightcmb }{\ensuremath{ 0.774 \pm 0.031}}
\newcommand{\wcdmomcmb }{\ensuremath{ 0.184 \pm 0.045}}
\newcommand{\wcdmwcmb }{\ensuremath{ -1.63 \pm 0.28}}
\newcommand{\wcdmsigcurrent }{\ensuremath{ 0.772 \pm 0.037}}
\newcommand{\wcdmwcurrent }{\ensuremath{ -1.07 \pm 0.20}}
\newcommand{\wcdmseightcurrent }{\ensuremath{ 0.743 \pm 0.048}}
\newcommand{\wcdmomcurrent }{\ensuremath{ 0.279 \pm 0.042}}
\newcommand{\wcdmaszcurrent }{\ensuremath{ 5.1 \pm 1.2}}
\newcommand{\wcdmbszcurrent }{\ensuremath{ 1.631 \pm 0.068}}
\newcommand{\wcdmcszcurrent }{\ensuremath{ 0.73 \pm 0.24}}
\newcommand{\wcdmdszcurrent }{\ensuremath{ 0.176 \pm 0.071}}
\newcommand{\wcdmwcurrentwcmb }{\ensuremath{ -1.30 \pm 0.10}}
\newcommand{\wcdmsignext }{\ensuremath{ 0.039}}
\newcommand{\wcdmsignextcmbcal }{\ensuremath{ 0.016}}
\newcommand{\wcdmsignextbothcal }{\ensuremath{ 0.014}}
\newcommand{\wcdmsigreductionnext }{\ensuremath{ 2.4}}
\newcommand{\wcdmwnext }{\ensuremath{ 0.19}}
\newcommand{\wcdmwnextcmbcal }{\ensuremath{ 0.15}}
\newcommand{\wcdmwnextbothcal }{\ensuremath{ 0.14}}
\newcommand{\wcdmwreductionnext }{\ensuremath{ 1.3}}
\newcommand{\wcdmseightnext }{\ensuremath{ 0.051}}
\newcommand{\wcdmseightnextcmbcal }{\ensuremath{ 0.025}}
\newcommand{\wcdmseightnextbothcal }{\ensuremath{ 0.023}}
\newcommand{\wcdmhnext }{\ensuremath{ 0.030}}
\newcommand{\wcdmhnextcmbcal }{\ensuremath{ 0.028}}
\newcommand{\wcdmhnextbothcal }{\ensuremath{ 0.024}}
\newcommand{\wcdmomnext }{\ensuremath{ 0.026}}
\newcommand{\wcdmomnextcmbcal }{\ensuremath{ 0.025}}
\newcommand{\wcdmomnextbothcal }{\ensuremath{ 0.024}}
\newcommand{\wcdmsigfuture }{\ensuremath{ 0.016}}
\newcommand{\wcdmsigfuturecmbcal }{\ensuremath{ 0.0044}}
\newcommand{\wcdmsigfuturebothcal }{\ensuremath{ 0.0040}}
\newcommand{\wcdmsigfuturebothcaloneperc }{\ensuremath{ 0.0046}}
\newcommand{\wcdmsigreductionfuture }{\ensuremath{ 3.6}}
\newcommand{\wcdmwfuture }{\ensuremath{ 0.028}}
\newcommand{\wcdmwfuturecmbcal }{\ensuremath{ 0.029}}
\newcommand{\wcdmwfuturebothcal }{\ensuremath{ 0.023}}
\newcommand{\wcdmwfuturebothcaloneperc }{\ensuremath{ 0.020}}
\newcommand{\wcdmwreductionlsst }{\ensuremath{ 1.2}}
\newcommand{\wcdmwreductionlsstoneperc }{\ensuremath{ 1.4}}
\newcommand{\wcdmseightfuture }{\ensuremath{ 0.016}}
\newcommand{\wcdmseightfuturecmbcal }{\ensuremath{ 0.0059}}
\newcommand{\wcdmseightfuturebothcal }{\ensuremath{ 0.0059}}
\newcommand{\wcdmseightfuturebothcaloneperc }{\ensuremath{ 0.0059}}
\newcommand{\wcdmhfuture }{\ensuremath{ 0.012}}
\newcommand{\wcdmhfuturecmbcal }{\ensuremath{ 0.0092}}
\newcommand{\wcdmhfuturebothcal }{\ensuremath{ 0.0071}}
\newcommand{\wcdmhfuturebothcaloneperc }{\ensuremath{ 0.0072}}
\newcommand{\wcdmomfuture }{\ensuremath{ 0.0063}}
\newcommand{\wcdmomfuturecmbcal }{\ensuremath{ 0.0057}}
\newcommand{\wcdmomfuturebothcal }{\ensuremath{ 0.0052}}
\newcommand{\wcdmomfuturebothcaloneperc }{\ensuremath{ 0.0050}}

\newcommand{\volwcdmnextcmb }{\ensuremath{ 2.8}}

\newcommand{\volwcdmnextboth}{\ensuremath{ 4.1}}

\newcommand{\volwcdmfutcmb }{\ensuremath{ 4.8}}

\newcommand{\volwcdmfutboth}{\ensuremath{ 6.1}}
\defcitealias{baxter15}{B15} 
\defcitealias{bocquet19}{B19}

\shorttitle{CMB-cluster lensing Cosmological constraints from SPT-SZ} \shortauthors{Chaubal et al.}
\begin{document} \title{Improving cosmological constraints from galaxy cluster number counts with CMB-cluster-lensing data: Results from the SPT-SZ survey and forecasts for the future}

\correspondingauthor{Prakrut S. Chaubal} \email{pchaubal@student.unimelb.edu.au}

%\AuthorCallLimit=3
\newcommand{\Missouri}{Department of Physics and Astronomy, University of Missouri-Kansas City, 5110 Rockhill Road, Kansas City, MO 64110, USA}
\newcommand{\ParisSaclay}{Universite Paris-Saclay, CNRS, Institut d’Astrophysique Spatiale, 91405, Orsay, France}
\newcommand{\TriesteINFN}{INFN-Sezione di Trieste, Trieste, Italy}
\newcommand{\TriesteIFPU}{IFPU - Institute for Fundamental Physics of the Universe, via Beirut 2, 34014 Trieste, Italy}
\newcommand{\TriesteINAF}{INAF-Osservatorio Astronomico di Trieste, via G. B. Tiepolo 11, I-34143 Trieste, Italy}
\newcommand{\TriesteUni}{Astronomy Unit, Department of Physics, University of Trieste, via Tiepolo 11, I-3413 Trieste, Italy}
\newcommand{\MIT}{Kavli Institute for Astrophysics and Space Research, Massachusetts Institute of Technology, 77 Massachusetts Avenue, Cambridge, MA 02139, USA}
\newcommand{\Durham}{Department of Physics, Durham University, South Road, Durham, DH1 3LE, UK}
\newcommand{\CSIRO}{CSIRO Astronomy and Space Science, PO Box 1130, Bentley WA 6102, Australia}
\newcommand{\McGill}{Department of Physics and McGill Space Institute, McGill University, Montreal, Quebec H3A 2T8, Canada}
\newcommand{\KIPAC}{Kavli Institute for Particle Astrophysics and Cosmology, Stanford University, 452 Lomita Mall, Stanford, CA 94305, USA}
\newcommand{\Stanford}{Dept. of Physics, Stanford University, 382 Via Pueblo Mall, Stanford, CA 94305}
\newcommand{\Davis}{Department of Physics, University of California, Davis, CA, USA 95616}
\newcommand{\Penn}{Center for Particle Cosmology, Department of Physics and Astronomy, University of Pennsylvania, Philadelphia, PA,  USA 19104} 
\newcommand{\KICPChicago}{Kavli Institute for Cosmological Physics, University of Chicago, Chicago, IL, USA 60637}
\newcommand{\PhysicsUChicago}{Department of Physics, University of Chicago, Chicago, IL, USA 60637}
\newcommand{\AAUChicago}{Department of Astronomy and Astrophysics, University of Chicago, Chicago, IL, USA 60637}
\newcommand{\FNAL}{Fermi National Accelerator Laboratory, MS209, P.O. Box 500, Batavia, IL 60510}
\newcommand{\ArgonneHEP}{High Energy Physics Division, Argonne National Laboratory, Argonne, IL, USA 60439}
\newcommand{\EFIChicago}{Enrico Fermi Institute, University of Chicago, Chicago, IL, USA 60637}
\newcommand{\SLAC}{SLAC National Accelerator Laboratory, 2575 Sand Hill Road, Menlo Park, CA 94025}
\newcommand{\Caltech}{California Institute of Technology, Pasadena, CA, USA 91125}
\newcommand{\Berkeley}{Department of Physics, University of California, Berkeley, CA, USA 94720}
\newcommand{\Cifar}{Canadian Institute for Advanced Research, CIFAR Program in Cosmology and Gravity, Toronto, ON, M5G 1Z8, Canada}
\newcommand{\Colorado}{Center for Astrophysics and Space Astronomy, Department of Astrophysical and Planetary Sciences, University of Colorado, Boulder, CO, 80309}
\newcommand{\ESO}{European Southern Observatory, Karl-Schwarzschild-Stra{\ss}e 2, 85748 Garching, Germany}
\newcommand{\Colphys}{Department of Physics, University of Colorado, Boulder, CO, 80309}
\newcommand{\Illast}{Astronomy Department, University of Illinois at Urbana-Champaign, 1002 W. Green Street, Urbana, IL 61801, USA}
\newcommand{\Illphys}{Department of Physics, University of Illinois Urbana-Champaign, 1110 W. Green Street, Urbana, IL 61801, USA}
\newcommand{\UChicago}{University of Chicago, Chicago, IL, USA 60637}
\newcommand{\LBNL}{Physics Division, Lawrence Berkeley National Laboratory, Berkeley, CA, USA 94720}
\newcommand{\Michigan}{Department of Physics, University of Michigan, Ann  Arbor, MI, USA 48109}
\newcommand{\Munich}{Faculty of Physics, Ludwig-Maximilians-Universit\"{a}t, 81679 M\"{u}nchen, Germany}
\newcommand{\ExcellenceCluster}{Excellence Cluster Universe, 85748 Garching, Germany}
\newcommand{\MPE}{Max-Planck-Institut f\"{u}r extraterrestrische Physik, 85748 Garching, Germany}
\newcommand{\Dunlap}{Dunlap Institute for Astronomy \& Astrophysics, University of Toronto, 50 St George St, Toronto, ON, M5S 3H4, Canada}
\newcommand{\Minnesota}{Department of Physics, University of Minnesota, Minneapolis, MN, USA 55455}
\newcommand{\Melbourne}{School of Physics, University of Melbourne, Parkville, VIC 3010, Australia}
\newcommand{\CaseWestern}{Physics Department, Center for Education and Research in Cosmology and Astrophysics, Case Western Reserve University,Cleveland, OH, USA 44106}
\newcommand{\ArtInstChicago}{Liberal Arts Department, School of the Art Institute of Chicago, Chicago, IL, USA 60603}
\newcommand{\JPL}{Jet Propulsion Laboratory, California Institute of Technology, Pasadena, CA 91109, USA}
\newcommand{\CfA}{Harvard-Smithsonian Center for Astrophysics, Cambridge, MA, USA 02138}
\newcommand{\UToronto}{Department of Astronomy \& Astrophysics, University of Toronto, 50 St George St, Toronto, ON, M5S 3H4, Canada}
\newcommand{\BCCP}{Berkeley Center for Cosmological Physics, Department of Physics, University of California, and Lawrence Berkeley National Labs, Berkeley, CA, USA 94720}

% Authors

%\collaboration{SPT Collaboration}
\author{P.~S.~Chaubal}
\affiliation{\Melbourne}

\author{ C.~L.~Reichardt}
\affiliation{\Melbourne}

\author{N.~Gupta}
\affiliation{\Melbourne}
\affiliation{\CSIRO}

% \author{S.~Raghunathan}
% \affiliation{}

\author{B.~Ansarinejad}
\affiliation{\Melbourne}
% \affiliation{\Durham}

\author{K.~Aylor}
\affiliation{\Davis}

\author{L.~Balkenhol}
\affiliation{\Melbourne}

\author{E.~J.~Baxter}
\affiliation{\Penn}
\affiliation{\KICPChicago}
\affiliation{\AAUChicago}  

\author{F.~Bianchini}
\affiliation{\Stanford}

\author{B.~A.~Benson}
\affiliation{\FNAL}
\affiliation{\KICPChicago}
\affiliation{\AAUChicago}

\author{L.~E.~Bleem}
\affiliation{\ArgonneHEP} 
\affiliation{\KICPChicago}

\author{S.~Bocquet}
\affiliation{\Munich}

\author{J.~E.~Carlstrom}
\affiliation{\KICPChicago}
\affiliation{\PhysicsUChicago}
\affiliation{\ArgonneHEP}
\affiliation{\AAUChicago}
\affiliation{\EFIChicago}

\author{ C.~L.~Chang}
\affiliation{\ArgonneHEP}
\affiliation{\KICPChicago}
\affiliation{\AAUChicago}

\author{ T.~M.~Crawford}
\affiliation{\KICPChicago}
\affiliation{\AAUChicago}

\author{ A.~T.~Crites}
\affiliation{\KICPChicago}
\affiliation{\AAUChicago}
\affiliation{\Caltech}

\author{ T.~de~Haan}
\affiliation{\McGill}
\affiliation{\Berkeley}

\author{ M.~A.~Dobbs}
\affiliation{\McGill}
\affiliation{\Cifar}

\author{ W.~B.~Everett}
\affiliation{\Colorado}

\author{B.~Floyd}
\affiliation{\Missouri}

\author{ E.~M.~George}
\affiliation{\Berkeley}
\affiliation{\ESO}

\author{ N.~W.~Halverson}
\affiliation{\Colorado}
\affiliation{\Colphys}

\author{ W.~L.~Holzapfel}
\affiliation{\Berkeley}

\author{ J.~D.~Hrubes}
\affiliation{\UChicago}

\author{ L.~Knox}
\affiliation{\Davis}

\author{ A.~T.~Lee}
\affiliation{\Berkeley}
\affiliation{\LBNL}

\author{ D.~Luong-Van}
\affiliation{\UChicago}

\author{ J.~J.~McMahon}
\affiliation{\Michigan}

\author{ S.~S.~Meyer}
\affiliation{\KICPChicago}
\affiliation{\AAUChicago}
\affiliation{\EFIChicago}
\affiliation{\PhysicsUChicago}

\author{ L.~M.~Mocanu}
\affiliation{\KICPChicago}
\affiliation{\AAUChicago}

\author{ J.~J.~Mohr}
\affiliation{\Munich}
\affiliation{\ExcellenceCluster}
\affiliation{\MPE}

\author{ T.~Natoli}
\affiliation{\KICPChicago}
\affiliation{\PhysicsUChicago}
\affiliation{\Dunlap}

\author{ S.~Padin}
\affiliation{\KICPChicago}
\affiliation{\AAUChicago}

\author{ C.~Pryke}
\affiliation{\Minnesota}

\author{ J.~E.~Ruhl}
\affiliation{\CaseWestern}

\author{ F. Ruppin}
\affiliation{\MIT}

\author{ L.~Salvati}
\affiliation{\TriesteINAF}
\affiliation{\TriesteIFPU}
\affiliation{\ParisSaclay}

\author{A.~Saro}
\affiliation{\TriesteUni}
\affiliation{\TriesteINAF}
\affiliation{\TriesteIFPU}
\affiliation{\TriesteINFN}

\author{ K.~K.~Schaffer}
\affiliation{\KICPChicago}
\affiliation{\EFIChicago}
\affiliation{\ArtInstChicago}

\author{ E.~Shirokoff}
\affiliation{\Berkeley} 
\affiliation{\KICPChicago} 
\affiliation{\AAUChicago}

\author{ Z.~Staniszewski}
\affiliation{\CaseWestern}
\affiliation{\JPL}

\author{ A.~A.~Stark}
\affiliation{\CfA}

\author{J.~D.~Vieira}
\affiliation{\Illast} 
\affiliation{\Illphys} 

\author{R.~Williamson}
\affiliation{\KICPChicago} 
\affiliation{\AAUChicago}

\begin{abstract} 
    We show the improvement to cosmological constraints from galaxy cluster surveys with the addition of CMB-cluster lensing data. 
    We explore  the cosmological implications of adding mass information from the 3.1\,$\sigma$ detection of gravitational lensing of the cosmic microwave background (CMB) by galaxy clusters
    to the Sunyaev-Zel'dovich (SZ) selected galaxy cluster sample from the 2500 deg$^2$  SPT-SZ survey and targeted optical and X-ray followup data.  
    In the \lcdm{} model, the combination of the cluster sample with the    \planck{} power spectrum measurements prefers $\sigma_8 \left(\Omega_m/0.3 \right)^{0.5} = \seightcurrentwcmb$. 
    Adding the cluster data reduces the uncertainty on this quantity  by a factor of \seightreductioncurrent, which is unchanged whether or not the 3.1\,$\sigma$ CMB-cluster lensing measurement is included. 
    We then forecast the impact of  CMB-cluster lensing measurements with future cluster catalogs.  
    Adding CMB-cluster lensing measurements to the SZ cluster catalog of
    the on-going SPT-3G survey is expected to improve the expected constraint on the dark energy equation of state $w$ by a factor of \wcdmwreductionnext{} to $\sigma(w) = \wcdmwnext$.
    We find the largest improvements from CMB-cluster lensing measurements to be for \sig{}, where adding CMB-cluster lensing data to the cluster number counts  reduces the expected uncertainty on \sig{} by factors of \wcdmsigreductionnext{} and \wcdmsigreductionfuture{} for SPT-3G and CMB-S4 respectively. 
\end{abstract}


\section{Introduction} 
\label{sec:intro} 
Galaxy clusters are  the largest gravitationally collapsed structures and a key testing ground of cosmological models
of structure growth \citep{allen11}.  The number density of galaxy clusters  depends sensitively upon cosmological parameters, particularly those that affect
late-time structure growth such as the sum of the neutrino masses, the dark energy equation of state, and matter density
\citep{wang98,haiman01,weller02,weller03,holder06,shimon11}.  Upcoming surveys such as eROSITA \citep{merloni12}, LSST \citep{lsst09, lsst18}
and CMB-S4 \citep{cmbs4collab19} are expected to detect tens of thousands of galaxy clusters at different wavelengths, and will dramatically improve the cosmological constraints from cluster
cosmology.

Galaxy clusters already yield interesting constraints on the matter density \Om{} and the amplitude of density fluctuations \sig{}
\citep{bocquet19,zubeldia19,to20}.  
The cosmological constraints are limited, however, by the uncertainty on the masses of galaxy clusters and can
be biased if the cluster mass-observable scaling relations are mis-estimated.  
Current cluster mass estimates are typically based on assuming a
power-law scaling relationship between observed quantities (such as the X-ray observable \yx{}) and cluster masses. 
Observationally expensive optical weak lensing measurements are used to normalize the scaling relation \cite[e.g.,][]{dietrich19}.  
These optical weak lensing mass measurements should substantially improve with surveys like LSST and Euclid \citep{lsst18, euclid19}.
At higher redshifts ($z \gtrsim 1$), optical weak lensing becomes increasingly difficult due to a dearth of background galaxies and difficulties in measuring their shape with blending and lower signal to noise. 
High-redshift mass information is important as there are suggestions that scaling relations calibrated at lower redshifts may mis-estimate the masses at higher redshifts \citep{zohren19, salvati18, salvati19}.

Galaxy clusters also gravitationally lens the cosmic microwave background (CMB), an effect referred to as CMB-cluster lensing and first considered by
\citet{seljak00b}.  
While useful as an independent cross-check on optical weak lensing cluster masses at low redshift, CMB-cluster lensing is particularly
useful at higher redshifts.  
Since all CMB photons originate at the same extremely high redshift, $z\simeq 1100$, the signal-to-noise of CMB-cluster lensing
does not drop as the cluster redshift increases \citep{melin15}.  
This also simplifies the measurement (and eliminates related uncertainties), as one does not need to calculate intrinsic alignments, boost factors, or the redshift distribution to background sources. 
The problem of estimating the masses of clusters from their
CMB lensing signals has been extensively considered \citep{seljak00b, holder04, vale04, dodelson04, lewis06, lewis06a, hu07,raghunathan17,raghunathan19b,gupta20b}.   
Actual measurements of the CMB-cluster lensing signal have followed as CMB surveys have advanced,  from the first detections in 2015 \citep{madhavacheril15, baxter15, planck15-24} to  $\sim$15\% mass measurements of
different cluster samples today \citep{baxter18, raghunathan19a}. 

In this work, we present the first cosmological analysis of the SPT-SZ galaxy cluster sample that includes CMB-cluster lensing information. 
The SPT-SZ survey detected galaxy clusters from the imprint of thermal SZ (tSZ) signatures on the background primary CMB anisotropies \citep{bleem15b}.
\citet[][hereafter B19]{bocquet19} presented cosmological constraints from this sample along with X-ray observations and optical weak-lensing measurements.  
We add the CMB-cluster lensing mass measurement of \citet[][hereafter B15]{baxter15} to that dataset, and look at the implications for the combined dataset on the
\LCDM{} and \wCDM{} cosmological models.  
We follow this by presenting forecasts for the cosmological constraints from future CMB-cluster lensing measurements with
SPT-3G \citep{benson14} and CMB-S4 \citep{cmbs4collab19}. 
We find that CMB-cluster lensing mass measurements substantially improve the predicted constraints on the dark energy equation of state parameter $w$
from future cluster catalogs.

The paper is organized as follows.  
In  \S\ref{sec:data}, we review the datasets used in this analysis.  
We describe the analysis methods in \S\ref{sec:method}.
In \S\ref{sec:results}, we present the cosmological constraints from the current CMB-cluster lensing measurement.  
In  \S\ref{sec:forecast}, we forecast the
constraints expected from the ongoing SPT-3G and future CMB-S4 surveys.  
Finally, we conclude in  \S\ref{sec:conclusions_and_outlook}. 
Throughout this work, we report galaxy cluster masses in terms of either $M_{\rm 200}$ or $M_{\rm 500}$, the mass contained within the radius where the mean density is 200 (500) times the critical density of the Universe.

\section{The Cluster catalog from the 2500d SPT-SZ survey} 
\label{sec:data}

The main dataset in this work is the galaxy cluster sample from the 2500d SPT-SZ survey \citep{bleem15b}, which provides a measure of the SZ detection significance
and redshift for each cluster in the sample.  As in the previous cosmological analysis by \citetalias{bocquet19}, we supplement the SZ cluster catalog with
follow-up X-ray and optical weak-lensing observations.  
The new addition in this work is that we add the $3.1\sigma{}$ CMB-cluster lensing mass measurement from \citetalias{baxter15} for a stack of 513 of galaxy clusters in the sample.
This sub-sample of 513 clusters is chosen by selecting only those clusters from the 2500d SPT-SZ catalog which have measured optical redshifts.
We refer to the combination of SPT number counts, X-ray and weak lensing follow up, and CMB cluster lensing datasets as SPT clusters. 
 We briefly describe these datasets in the following subsections. 

For some parameter fits, we also include measurements of the CMB TT, TE and EE power spectra from the 2018 data release of the \planck{} satellite
\citep{planck18-5}.  
We refer to this dataset as `Planck' throughout rest of the work.  
The \planck{} CMB data allow us to demonstrate where  clusters and CMB-cluster lensing add the most information.

\subsection{SZ detection significance and cluster redshift} 
\label{sub:spt_sz_data}

The SZ detection significance and cluster redshift (or lower limit on redshift) are reported for all cluster candidates in the \citet{bleem15b} catalog and were
later updated in \citetalias{bocquet19}.  The reported significance is the maximum across a set of matched filters (to allow for variations in the cluster
angular radius with redshift and mass), and therefore is biased high on average.  
To avoid this biasing in the mass estimates, we follow \citetalias{bocquet19} in using the unbiased significance $\zeta =\sqrt{(\xi^2 -3)}$ as a mass proxy. 
A detailed discussion on the validity of this approach can be found in \citet{vanderlinde10}.  
As in \citetalias{bocquet19}, we model the relationship between the unbiased significance $\zeta$ and cluster mass
$M_\mathrm{500}$ as: 
\begin{align} 
    \zeta = \ &\asz \left( \frac{M_{500}h_{70}}{4.3 \times 10^{14}  M_{\odot}}\right)^{B_\mathrm{SZ}} \left( \frac{E\left(z\right)}{E\left(0.6\right)}\right)^{C_{\mathrm{SZ}}} \ , \label{zeta_mass} 
\end{align} 
where \asz, \bsz, and  \csz  are free parameters in the model fits (see Table \ref{tab:priors}) and $E(z)$ is the dimensionless Hubble parameter.  Here $h_{70}$ is
the Hubble constant divided by 70 km s$^{-1}$ Mpc$^{-1}$, and $z$ is the cluster redshift. The intrinsic scatter in $\mathrm{ln}\zeta$ at a fixed mass and
redshift, is modeled as a Gaussian scatter with width $\sigmalnzeta$ and is also left as a free parameter of the model.

\subsection{Weak-lensing shear profiles}

Thirty-two clusters have optical weak lensing shear profiles, with 13 from the Hubble Space Telescope and 19 from ground-based Megacam/Magellan imaging \citep{schrabback18,
dietrich19}.  The shear profiles of these clusters are compared to the expected weak lensing shear profiles under the assumption of a Navarro-Frenk-White (NFW) profile \citep{navarro97} for the cluster density.  
We allow for a systematic bias $b_{WL}$ between the halo mass $M_{\rm halo}$ and inferred lensing mass $M_{WL}$, \begin{equation} M_{WL} = b_{WL} M_{\rm halo} \ .  \end{equation} 
We refer the reader to Eqn.~9 in \citetalias{bocquet19} for the breakdown of $b_{WL}$ into different sources of uncertainty in the weak lensing observations. 
The priors on these uncertainties are included in Table~\ref{tab:priors} under the \emph{WL modeling} section. 
The weak-lensing  model is described in more detail by \citetalias{bocquet19}.

\subsection{X-ray \yx{} data}

As in \citetalias{bocquet19}, we use X-ray observations of 89 galaxy clusters taken through a \chandra{} X-ray visionary project \citep{mcdonald13,mcdonald17}.
The X-ray data is used to estimate \yx{} (the product of the gas mass and X-ray temperature) within $r_{500}$ for each cluster.  We assume a scaling relation
between \yx{} and the cluster mass $M_{500}$  of the form:

\begin{align} 
    \mathrm{ln}\left( \frac{M_{500}h_{70}}{8.37\times 10^{13}M_{\odot}} \right) = \ &\mathrm{ln} A_{\yx} + B_{\yx} \langle \mathrm{ln} \yx \rangle \\ \nonumber 
                                                                       & + B_{\yx} \mathrm{ln}\ \left( \frac{h_{70}^{5/2}}{3\times 10^{14} M_{\odot}\mathrm{keV}} \right) \\ \nonumber 
                                                                       & + C_{\yx}\mathrm{ln} \ E\left(z\right) \ .
\end{align} 

The intrinsic scatter in $\mathrm{ln}\,\yx$ at fixed mass and redshift is modeled as a normal distribution with width  $\sigmalnyx$.

\subsection{CMB-Cluster lensing measurement} 
\label{sub:cmblensing}
CMB photons are deflected by the gravitational pull of galaxy clusters.  
This deflection remaps the CMB anisotropy, and introduces a dipole-like signal
aligned with the local gradient in the primary CMB anisotropy \citep{lewis06}.  
\citetalias{baxter15}  extracted this CMB-cluster lensing signal from the SPT-SZ survey data at the positions of clusters in the SPT-SZ sample.  
To avoid being biased by the cluster's own tSZ signal, \citetalias{baxter15} used a linear
combination of the 90, 150 and 220 GHz maps from the SPT-SZ survey to make a tSZ-free map for the analysis. 
We refer the reader to \citetalias{baxter15} for further details on the measurement.

For the SPT-SZ catalog sub-sample described in \S\ref{sec:data}, \citetalias{baxter15} found the mean mass of the stacked clusters to be $\bar{M}_{200} = (5.1 \pm
2.1)\times 10^{14} M_{\odot}$. 
We convert $M_{200}$ to $M_{\rm 500}$  by
assuming a concentration parameter $c=3$ and the same flat \LCDM{} cosmological parameters used in \citetalias{baxter15} ($\Om{} = 0.3$,  $h=0.7$) for the redshift of $z=0.7$.
This gives us a value of $M_{\rm 500} = \left( 3.49 \pm 0.74  \right) \times 10^{14} M_{\odot} $ which we use in our analysis.
We note that converting the mean mass of the stack from $M_{200}$ to $M_{500}$ is not equivalent to converting individual cluster masses before stacking as the concentration-mass 
relation is redshift dependent. 
For this sample, this approximation results in a $\sim$ 2\% systematic error, which is negligible at the current statistical uncertainty, although the approximation may be inadequate for future high-S/N mass measurements.

\section{Likelihood} 
\label{sec:method}

As in past SPT-SZ cluster analyses \citep{reichardt13, dehaan16, bocquet19}, we derive cosmological constraints from galaxy clusters by using the Cash statistic
\citep{cash79} to compare the expected number of clusters with the observed number as a function of the SZ signal and redshift.  The number density of
clusters is predicted from the matter power spectrum and mass-observable scaling relations for each set of model parameters.  We briefly review the
likelihood\footnote{ \href{https://github.com/SebastianBocquet/SPT\_SZ\_cluster\_likelihood}{https://github.com/SebastianBocquet/SPT\_SZ\_cluster\_likelihood}}
here, which is presented in more detail by \citetalias{bocquet19}, before describing how we incorporate the new CMB-cluster lensing information. 

We choose to express the  likelihood function in three parts: cluster abundances ($\mathcal{L}_\mathrm{abund}$),  mass calibration from the weak lensing and X-ray observations ($\mathcal{L}_\mathrm{fol}$), and mass calibration from the CMB-cluster lensing observation ($\mathcal{L}_\mathrm{CL}$). 
The abundance part (which is unchanged from \citetalias{bocquet19})  calculates the chance of finding a catalog of clusters with the specified redshifts and SZ significances as a function of
the cosmology and scaling relations.  
As in \citetalias{bocquet19}, the X-ray and weak-lensing mass calibration likelihood is expressed as:
\begin{equation} 
    \label{eq:masscalib} 
    \begin{split} 
        \mathcal{L}_\mathrm{fol} \equiv P(\yx^\mathrm{obs}, & g_\mathrm{t}^\mathrm{obs} | \xi, z, \vec p) = \\ 
                                                           & \iiiint dM\, d\zeta\, d\yx\, dM_\mathrm{WL}\, \left [ \right. \\ 
                                                           & P(\yx^\mathrm{obs}|\yx) P(g_\mathrm{t}^\mathrm{obs}|M_\mathrm{WL}) P(\xi|\zeta) \\ 
                                                           & P(\zeta,\yx,M_\mathrm{WL}|M,z,\vec p) P(M|z,\vec p) \left. \right ] \ .
    \end{split} 
\end{equation}
This equation gives the likelihood of observing the follow-up X-ray, $\yx^\mathrm{obs}$, and weak lensing, $g_\mathrm{t}^\mathrm{obs}$, observables for a cluster
detected with SZ significance $\xi$. 
Here, $\vec{p}$ represents cosmological and scaling relation parameters. 
We assume the systematics in the CMB-cluster lensing
measurement to be uncorrelated with other observations.  
The notation adopted for other variables is identical to that of \citetalias{bocquet19}.

While we could exactly mirror the approach used for including weak lensing data, the CMB-cluster lensing signal from individual clusters is too weak to
justify the computational complexity.  
Instead, we take the observed mean mass from  CMB-cluster lensing $\bar{M}_{200} = (5.1 \pm 2.1)\times 10^{14} M_{\odot}$ as a prior on the 
modeled mean mass of the sample,  $\bar{M}$: 
\begin{equation} 
\bar{M} = \frac{1}{N}\sum_i \iint d\xi_i dz_i \ M_i P(M_i | \xi_i, z_i) P(\xi_i,z_i | \vec{p}) \ .  
\end{equation}
Given the number of clusters in the sample, we approximate the integral by taking the mass at the peak of the posterior for each cluster in the sample.

\begin{deluxetable}{ll} 
    \tablecaption{ Parameter priors \label{tab:priors} } 
    \tablehead{\colhead{Parameter} & \colhead{Prior}}
    \startdata
\multicolumn{2}{l}{SZ scaling relation}\\ \asz & $\mathcal U(1,10)$ \\ \bsz & $\mathcal U(1,2.0)$ \\ \csz & $\mathcal U(-1,2)$ \\ \sigmalnzeta & $\mathcal
U(0.01,2.0)$ \\ 
\tableline\tableline 
\multicolumn{2}{l}{Priors for the SPT-SZ cluster catalog}\\
\multicolumn{2}{l}{~~~X-ray \yx\ scaling relation}\\ \ayx & $\mathcal U(3, 10)$ \\ \byx & $\mathcal U(0.3,0.9)$ \\ \cyx & $\mathcal
U(-1,0.5)$ \\ \sigmalnyx & $\mathcal U(0.01,0.5)$\\ $d\ln M_\mathrm{g}/d\ln r$ & $\mathcal{N}(1.12, 0.23^2)$\\ 
\tableline \multicolumn{2}{l}{~~~WL modeling}\\
$\delta_\mathrm{WL,bias}$ & $\mathcal{N}(0,1)$ \\ $\delta_\mathrm{Megacam}$ & $\mathcal{N}(0,1)$ \\ $\delta_\mathrm{HST}$ & $\mathcal{N}(0,1)$ \\
$\delta_\mathrm{WL,scatter}$ & $\mathcal{N}(0,1)$ \\ $\delta_{\mathrm{WL,LSS}_\mathrm{Megacam}}$ & $\mathcal{N}(0,1)$ \\ $\delta_{\mathrm{WL,LSS}_\mathrm{HST}}$
                             & $\mathcal{N}(0,1)$ \\ 
                             \tableline \multicolumn{2}{l}{~~~Correlated scatter}\\ $\rho_\mathrm{SZ-WL}$ & $\mathcal U(-1,1)$ \\
$\rho_\mathrm{SZ-X}$ & $\mathcal U(-1,1)$ \\ $\rho_\mathrm{X-WL}$ & $\mathcal U(-1,1)$ \\ &$\det (\vec\Sigma_\text{multi-obs}) >0$ \\ 
\tableline
\tableline \multicolumn{2}{l}{Priors on cluster-only chains} \\
$\Omega_b h^2$ & $\mathcal{N}(0.02212,0.00022^2)$ \\
$\tau$ & $\mathcal{N}(0.0522,0.0080^2)$\\
$10^9A_s$ & $\mathcal{N}(2.092 , 0.033^2)$ \\
$n_s$ & $\mathcal{N}(0.9626, 0.0057^2)$ \\
\enddata
\tablecomments{
    The parameter priors used in this analysis are listed here. 
    The symbol $\mathcal{U}$ denotes a uniform prior over the given range while  
    $\mathcal{N}(\mu,\sigma^2)$ denotes a Gaussian prior centered at $\mu$ with variance $\sigma^2$. 
    The SZ scaling relation priors are used for all results in this work that include cluster data, while the cluster-only priors listed in the bottom section are only used in cluster-only-MCMCs. 
    The priors in the X-ray, WL modeling and Correlated scatter section are used for the SPT-SZ cluster data, but not in forecasts for future experiments. 
}
\end{deluxetable}

\section{Parameter constraints}
\label{sec:results}

We now turn to the cosmological implications of the CMB-cluster lensing measurement and cluster catalog described in  \S\ref{sec:data} using the likelihood function
described in \S\ref{sec:method}. 
All MCMC analyses use the same priors for the scaling relations, which are listed in Table \ref{tab:priors}.

We infer cosmological constraints using the publicly available COSMOSIS  parameter estimation code \citep{zuntz14}, running the Boltzmann code package
CAMB \citep{lewis99}.  
We use the \emph{Multinest} or \emph{emcee} samplers \citep{feroz08,foreman13} as implemented by COSMOSIS. 
Multinest is run with 250 live points with a tolerance value of 0.1.
We look at two cosmological models: the standard six-parameter \lcdm{} model with fixed $\sumMnu=0.06$\,eV, and a well-motivated extension to \lcdm{} where the
dark energy equation of state, $w$, is allowed to vary.

\subsection{\lcdm{} Cosmology} \label{subsec:lcdmcosmo} Galaxy cluster number counts are very sensitive to the growth of matter perturbations.  
Previous works
have found galaxy clusters constrain best the parameter combination $S_8 = \sigma_8 \left(\Omega_m/0.3\right)^{0.5}$.  
We find for the SPT cluster sample with \planck{} power
spectrum measurement: 

\begin{equation} 
    S_8 = \seightcurrentwcmb \ .
\end{equation} 

The uncertainty is larger than \planck-only by a factor of  \seightreductioncurrent{} , due to the tension between the \planck{} data favoring $S_8 = \seightcmb$ and cluster data favoring a lower $S_8 = \seightcurrent$.
The result is similar to what was found in \citetalias{bocquet19} so we do not attribute it to CMB-cluster lensing.
The similarity is understandable since the S/N on the CMB-cluster lensing is low compared to optical weak-lensing. For instance,
changing the mass normalization \asz{} from 4.4 to 5.5, the weak-lensing log-likelihood changes by $\Delta \rm{ln} \mathcal{L}_{WL} = -5.8$,
15 times greater than the change in the CMB-cluster lensing log-likelihood of $\Delta \rm{ln} \mathcal{L}_{CMBcl} = -0.38$ for the same shift. 
As noted above for $S_8$, the modest tension between the cluster and \planck{} data leads to slightly wider constraints for the combined dataset on  $\Om$ and $\sig$:
\begin{eqnarray} 
    \Om & =& \omcurrentwcmb \ , \\ 
    \sig & =& \sigcurrentwcmb\ .  
\end{eqnarray}
We report the parameter constraints on selected cosmological and scaling relation parameters in Table \ref{tab:constraints}. 

\begin{figure}[htp] \centering \includegraphics[scale=1]{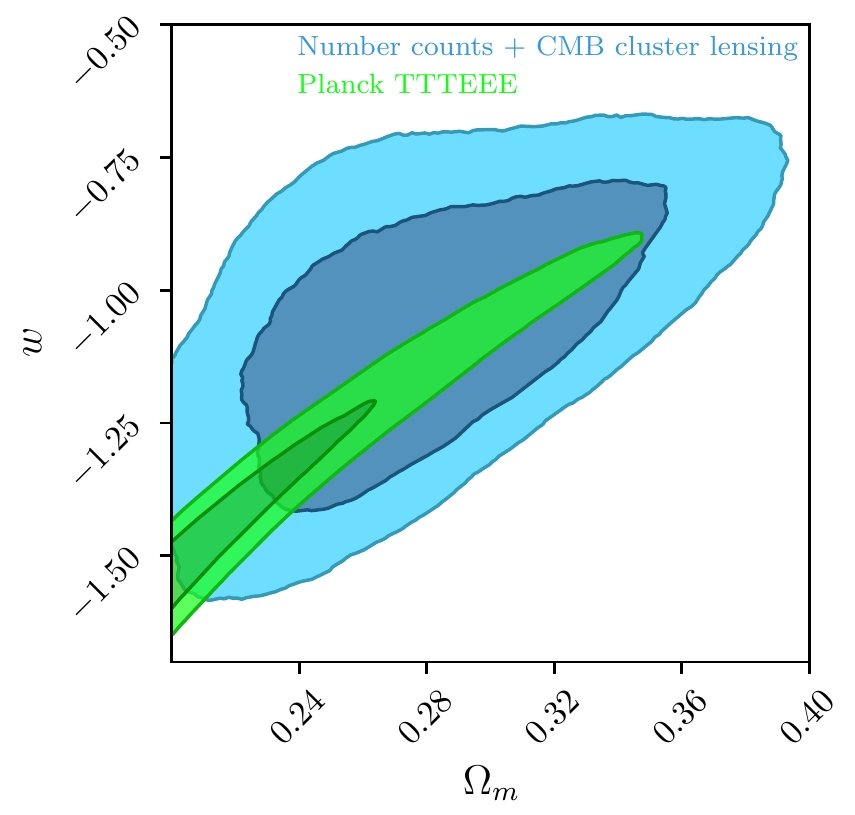} \caption{ Constraints on \Om{} and $w$ in the \wCDM{} model from the SPT-SZ cluster dataset (blue contours) and the \planck{} TTTEEE power spectra (green contours). 
The SPT-SZ cluster count constraints are obtained using CMB-cluster lensing information along with information from follow-up datasets. 
The cluster data help break the degeneracy between  \Om{} and $w$  that exists in the CMB power spectra alone. 
} 
\label{fig:sptsz} \end{figure}

\begin{deluxetable*}{lcccc}
\tablecaption{ 
\label{tab:constraints}
Parameter Constraints for the \planck{} and SPT-SZ surveys
}

\tablehead{ \colhead{Parameter}  & \multicolumn{2}{c}{\lcdm}  & \multicolumn{2}{c}{\wCDM}    }
\startdata
                                 & \planck{} & SPT Clusters & \planck{} & SPT Clusters \\
\tableline
\Om          & \omcmb{}& \omcurrent{} & \wcdmomcmb{} & \wcdmomcurrent{}  \\ 
\sig         & \sigcmb{} & \sigcurrent{} & \wcdmsigcmb{} & \wcdmsigcurrent{}  \\ 
\seight      & \seightcmb{} & \seightcurrent{} & \wcdmseightcmb{} & \wcdmseightcurrent{} \\      
$w$          & -- & --                                        & \wcdmwcmb{} & \wcdmwcurrent{} \\ 
\tableline
\asz         & -- &\aszcurrent{}  & -- & \wcdmaszcurrent{}  \\ 
\bsz         & -- & \bszcurrent{}  & -- & \wcdmbszcurrent{}  \\ 
\csz         & -- & \cszcurrent{}  & -- & \wcdmcszcurrent{} \\ 
\sigmalnzeta & -- & \dszcurrent{}  & -- & \wcdmdszcurrent{}  \\ 
\tableline
\enddata
\tablecomments{
Summary of constraints obtained from including cluster data in our analysis for \lcdm{} and \wCDM{}  cosmological models. 
Constraints obtained from using \planck{} only dataset are given for comparison. 
}
\end{deluxetable*}

\subsection{\wCDM }
Clusters are an important probe of the late time Universe when dark energy dominates the energy budget.
We therefore consider the impact of the cluster abundance and CMB-cluster lensing measurement on the dark energy equation of state parameter $w$.  
The cluster data favors \begin{equation} w = \wcdmwcurrent\ ,  \end{equation}
consistent with a cosmological constant. 
As shown in Fig.~\ref{fig:sptsz}, the cluster abundance data prefers a higher value of the dark energy equation of state as the  matter density increases.
The detection significance of the \citetalias{baxter15} CMB-cluster lensing measurement is as yet too low to significantly tighten the allowed parameter volume. 
While this uncertainty on $w$ is modestly tighter than that inferred from \planck{} power spectra alone ($w = -1.56^{+0.19}_{-0.39}$), combining the cluster abundance and \planck{} CMB data significantly reduces the allowed region to:
 \begin{equation} w = \wcdmwcurrentwcmb\ .  \end{equation}

\section{Forecasts} 
\label{sec:forecast}

We now examine the expected impact of CMB-cluster lensing on the cosmological constraints from upcoming galaxy cluster surveys.  
Using the likelihood framework from
\S\ref{sec:method}, we forecast the results from two surveys: the on-going SPT-3G survey, and the planned CMB-S4 survey.  
We assume that SPT-3G will survey
1500\,\sqdeg{} with a temperature map noise level of 2.5\,\ukarcmin{} (polarization map noise level a factor of $\sqrt{2}$ higher) at 150\,GHz \citep{sobrin21} and produce a catalog of  $\sim$3600 clusters above a signal-to-noise of 4.5.  
After
galactic cuts, we assume the CMB-S4 survey will cover 60\% of the sky with a map noise level of 1.0\,\ukarcmin{} (polarization map noise level a factor of $\sqrt{2}$ higher) at 150\,GHz \citep{cmbs4collab19} and produce a catalog of
$\sim$135,000 clusters above a signal-to-noise of 4.5.  
CMB-S4 will survey 3\% of the sky to even lower noise levels, which is expected to add a further 17,000
clusters.  
Catalogs from both CMB-S4 surveys are used in the forecasts in this work. 
We look at the results for the cluster abundances alone, and in combination with mass information from optical weak lensing or CMB-cluster lensing. 
The redshift bins and the uncertainties for SPT-3G 
and CMB-S4 surveys are described below.

For the full SPT-3G survey, we expect CMB-cluster lensing to lead to a 4.6\% mass measurement across the entire cluster sample \citep{raghunathan17}.  Given the
high detection significance, we choose to subdivide the cluster catalog into four redshift bins to better constrain any redshift evolution in the relationship
between SZ flux and mass.  The four redshift bins are $[0.25, 0.55), [0.55,0.78), [0.78, 1.06)$, and $[1.06, 2.]$, which achieves a roughly equal number of clusters
and lensing detection significance in each bin.  The uncertainty on the average mass of the clusters in each of the four bins is taken to be 9.2\%. For simplicity, 
we assume equal constraining power in each of the bins.  
We do not
include the effect of systematic uncertainties, such as from tSZ contamination or errors in the assumed mass profile, but point interested readers towards
\cite{raghunathan17} for a discussion of potential systematic errors and their magnitude.  
The potential systematic biases are expected to be correctable to better than the mass uncertainties assumed in this work. 
We conservatively assume a 5\% mass calibration from optical weak lensing at
$z<0.8$, again implemented as four 10\% mass constraints on redshift bins running $[0.25,0.39),[0.39,0.53),[0.53,.67),$ and $[0.67,0.8]$, 
such as might be achieved from the final DES results \citep{mcclintock19}.

The CMB-S4 survey is expected to start in the second half of this decade.  As such, we assume substantially improved optical weak lensing mass measurements will
be available from, for instance, LSST or Euclid, and provide either a 2\% (conservative) or 1\% (goal) mass calibration \citep{grandis18}.  
As before, we implement this as either a 4\% or 2\% mass calibration in each of four
redshift bins that cover the redshift range from z = 0.25 to 0.8.  The lower noise CMB maps will also enable tighter mass constraints from CMB-cluster lensing.
From \citet{raghunathan17}, we estimate that the CMB-S4 wide survey will yield a 3\% mass calibration in each of the four redshift bins, while the deep survey
will yield a weaker (due to fewer clusters) 5\% mass calibration  in each redshift bin.  As with SPT-3G, we do not include the effect of systematic errors. 

\begin{figure*}[ht]
    \centering \includegraphics[scale=0.52]{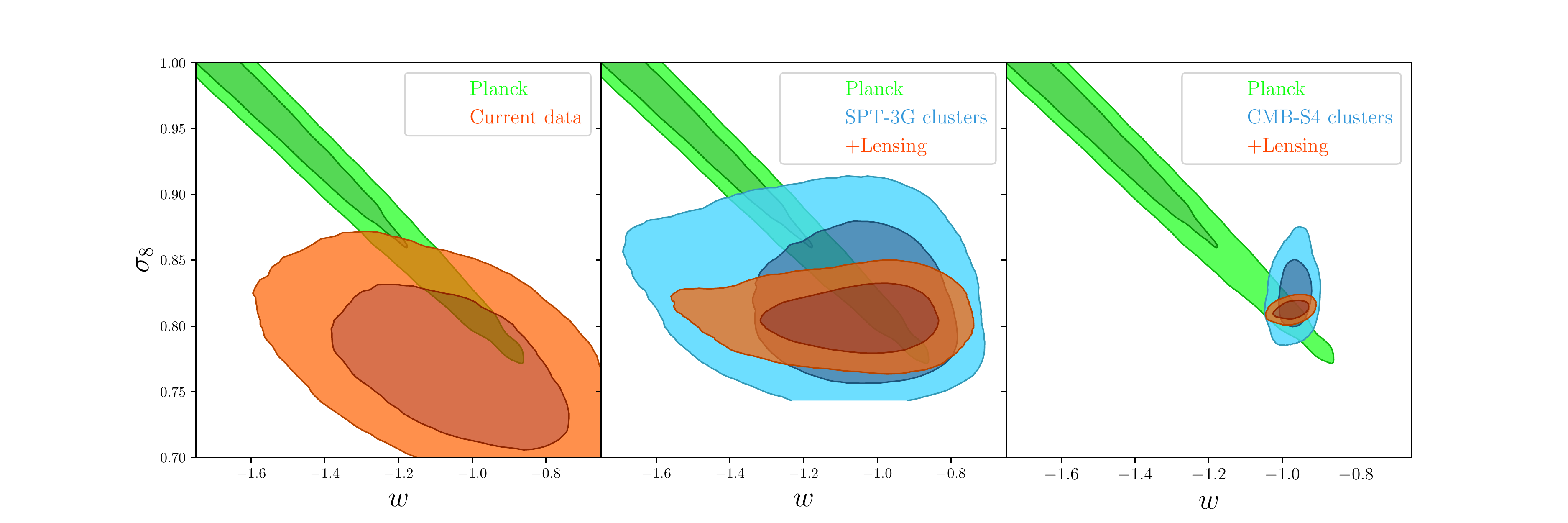} 

    \caption{ The 1 and 2\,$\sigma$ contours for $\sigma_8$ and $w$ in the \wCDM{} model for the SPT-SZ (left panel), SPT-3G (middle panel) and CMB-S4 (right panel) surveys. 
    The SPT-3G and CMB-S4 contours are forecasts from simulated cluster catalogs created for $\sigma_8 = 0.8126$ and $w = -1$.
    Parameter posterior distributions from the \planck{} CMB data are shown in green, while the posteriors from cluster number counts are shown in blue. 
    The posterior distributions from cluster number counts and CMB-cluster lensing are shown in orange.
        Adding CMB-cluster lensing information significantly improves the constraints on equation of dark energy parameter $w$ and \sig. } 
    \label{fig:forecasts} 
\end{figure*}

As shown in Table \ref{tab:forecasts} and Fig.~\ref{fig:forecasts}, we find that adding the mass information from optical weak lensing and CMB-cluster lensing
substantially improves cosmological constraints from galaxy cluster abundances with SPT-3G and CMB-S4.
 Assuming that that posteriors are approximately Gaussian, 
we calculate the allowed parameter volume as the square root of the determinant of the covariance matrix. 
The allowed parameter volume from the  cluster abundance data for the 7 parameters of the $w$CDM model  is reduced by a factor of \volwcdmnextboth{} for SPT-3G and \volwcdmfutboth{} for CMB-S4 by adding the CMB-cluster and optical lensing measurements.  
While
the absolute mass calibration is similar between the optical and CMB lensing channels ($\sim$\,5\% for SPT-3G and $\sim$\,2-3\% for CMB-S4), the higher redshift
lever arm in the CMB-cluster lensing measurement has advantages for the SZ cluster catalogs with their high median redshifts ($\sim$\,0.8 for both the SPT-3G
and CMB-S4 surveys).  
For the SPT-3G cluster sample, adding only the CMB-cluster lensing measurement reduces the parameter 
volume by a factor of  \volwcdmnextcmb. 
Adding both CMB-cluster lensing and optical weak lensing
improves the parameter volume by a factor of  \volwcdmnextboth, as stated above. 
 This translates to an improvement on $w$ from $\sigma(w) = \wcdmwnext$ for 
cluster counts to $\sigma(w) =  \wcdmwnextcmbcal$ with CMB-cluster lensing and $\sigma(w) =  \wcdmwnextbothcal$ with CMB-cluster lensing and optical weak lensing information (the latter two uncertainties are consistent given the number of samples in the MCMC). 
The expected constraint on $\sigma_8$ shows an even larger improvement, tightening from $\sigma(\sigma_8) = \wcdmsignext$ for 
cluster counts to $\sigma(\sigma_8) =  \wcdmsignextcmbcal$ with CMB-cluster lensing and $\sigma(\sigma_8) =  \wcdmsignextbothcal$ with CMB-cluster lensing and optical weak lensing information.  
The story is similar for CMB-S4. %, although  the optical weak lensing data from LSST is more significant. 
The 7-parameter volume is reduced by a factor of \volwcdmfutcmb{} (\volwcdmfutboth) by adding CMB-cluster lensing (both CMB-cluster lensing and a 2\% optical weak lensing measurement). 
Adding both the optical weak lensing and CMB-cluster lensing information brings $\sigma(w) = \wcdmwfuture$  down to $\sigma(w) = \wcdmwfuturebothcal$ for a 2\% mass calibration ($\sigma(w) = \wcdmwfuturebothcaloneperc$ for a 1\% mass calibration), a factor of \wcdmwreductionlsst{} (\wcdmwreductionlsstoneperc)
improvement over the cluster counts alone.  
The CMB-cluster lensing information substantially improves the constraint on $\sigma_8$ from the CMB-S4 cluster catalog by more than a factor of three, from $\sigma(\sigma_8) =  \wcdmsigfuture$ to $\sigma(\sigma_8) =  \wcdmsigfuturecmbcal$. 
Adding a 1\% (2\%) optical weak lensing mass measurement yields consistent results (within the sampling error) of $\sigma(\sigma_8) =  \wcdmsigfuturebothcaloneperc{}~(\wcdmsigfuturebothcal)$.
  CMB-cluster lensing
cluster mass measurements will be important to achieve the full potential of cluster cosmology over this decade.

\begin{deluxetable*}{llcccccc}
\tablecaption{ \label{tab:forecasts} Forecasts for Parameter Constraints for Upcoming Surveys }

\tablehead{ 
    \colhead{Survey} & \colhead{Data}  & \colhead{ $\Omega_m$ }  & \colhead{ $h$ }   & \colhead{ $w$ }  & \colhead{ $\sigma_8$ }  & \colhead{$S_8$}  }
\startdata
\tableline \tableline
    \planck{}           & CMB TTTEEE power spectra &        0.045  &     0.10   &     0.28   & 0.077 & 0.031      \\
\tableline
\multirow{4}{*}{SPT-3G} & Number counts         &\wcdmomnext & \wcdmhnext &\wcdmwnext & \wcdmsignext&  \wcdmseightnext \\ 
                        & ~~~ + CMB-cluster lensing           & \wcdmomnextcmbcal & \wcdmhnextcmbcal & \wcdmwnextcmbcal& \wcdmsignextcmbcal& \wcdmseightnextcmbcal  \\
                        & ~~~ + CMB-cluster and optical weak lensing  & \wcdmomnextbothcal & \wcdmhnextbothcal & \wcdmwnextbothcal& \wcdmsignextbothcal&  \wcdmseightnextbothcal \\
\tableline
\multirow{4}{*}{CMB-S4} & Number counts                & \wcdmomfuture & \wcdmhfuture &\wcdmwfuture & \wcdmsigfuture&  \wcdmseightfuture \\ 
                        & ~~~ + CMB-cluster lensing           & \wcdmomfuturecmbcal & \wcdmhfuturecmbcal & \wcdmwfuturecmbcal& \wcdmsigfuturecmbcal& \wcdmseightfuturecmbcal  \\
                        & ~~~ + CMB-cluster and 2\% optical weak lensing  & \wcdmomfuturebothcal & \wcdmhfuturebothcal & \wcdmwfuturebothcal& \wcdmsigfuturebothcal&  \wcdmseightfuturebothcal \\
                                                & ~~~ + CMB-cluster and 1\% optical weak lensing  & \wcdmomfuturebothcaloneperc & \wcdmhfuturebothcaloneperc & \wcdmwfuturebothcaloneperc& \wcdmsigfuturebothcaloneperc&  \wcdmseightfuturebothcaloneperc \\
\enddata
\tablecomments{
Cluster counts from SPT-3G and CMB-S4 can significantly improve cosmological constraints. 
We report here forecasted constraints in the 7-parameter \wCDM{} model for $w$, $\sigma_8$, and $S_8 \equiv \sigma_8 \sqrt{\Omega_m/0.3}$.
The second row has current uncertainties from the \planck{} 2018 TTTEEE data shown for comparison. 
The third through fifth rows have, in order, the expected uncertainties with the SPT-3G cluster counts , with the SPT-3G  cluster counts and CMB-cluster lensing mass measurement, with the SPT-3G  cluster counts and a DES-like optical weak lensing mass measurement, and with 
both the optical and CMB-cluster lensing mass measurements. 
The sixth through ninth rows are the same except for CMB-S4 and two options for an LSST-like optical survey that yields either a 1\% or 2\% mass measurement. 
Adding the optical weak lensing mass measurements to the CMB-S4 catalog does not improve estimates of large scale structure today (i.e.~$\sigma_8$) but does noticeably improve the constraints on the dark energy equation of state. 
}
\end{deluxetable*}

\section{Conclusions and Outlook}
\label{sec:conclusions_and_outlook}

We present the first cosmological parameter constraints incorporating CMB-cluster lensing mass estimates from the South Pole Telescope.  While the CMB-cluster
lensing mass information does not yet substantively improve cosmological constraints as compared to \citetalias{bocquet19}, this work serves
as a demonstration for the method which will be important for the next generation of large galaxy cluster surveys. 

We show that adding CMB-cluster lensing mass measurements should  significantly improve cosmological constraints from on-going cluster surveys such as SPT-3G.
In the 7-parameter $w$CDM cosmological model, we find that adding CMB-cluster lensing mass estimates to cluster number counts leads to a factor of \wcdmwreductionnext{} reduction
in the uncertainty of $w$ and a factor of \wcdmsigreductionnext{} on $\sigma_8$. 

CMB-cluster lensing data remains significant for the larger galaxy cluster catalog expected for CMB-S4.  
For CMB-S4, we find the CMB-cluster lensing data reduces the uncertainty on $\sigma_8$ by a factor of \wcdmsigreductionfuture. 
CMB-cluster lensing has the potential to significantly expand the cosmological information we can extract
from galaxy cluster surveys. 

\acknowledgements{
The South Pole Telescope program is supported by the National Science Foundation (NSF) through award OPP-1852617. 
Argonne National Laboratory's work was supported by the U.S. Department of Energy, Office of High Energy Physics, under contract DE-AC02-06CH11357. 
We also acknowledge support from the Argonne Center for Nanoscale Materials.  
The Melbourne group acknowledges support from the Australian Research Council's Discovery Projects scheme (DP200101068).
AAS acknowledges support by U.S. National Science Foundation grant AST-1814719. 
AS is supported by the FARE-MIUR grant ’ClustersXEuclid’ R165SBKTMA, INFN InDark, and by the ERC-StG ‘ClustersXCosmo’  grant agreement 716762.
The data analysis pipeline also uses the scientific python stack \citep{hunter07, jones01, vanDerWalt11}.
We acknowledge the use of the Spartan, a high performance computing facility at the University of Melbourne \citep{spartan}.
}

\bibliography{spt}{} \bibliographystyle{aasjournal}

\begin{thebibliography}{}
\expandafter\ifx\csname natexlab\endcsname\relax\def\natexlab#1{#1}\fi
\providecommand{\url}[1]{\href{#1}{#1}}
\providecommand{\dodoi}[1]{doi:~\href{http://doi.org/#1}{\nolinkurl{#1}}}
\providecommand{\doeprint}[1]{\href{http://ascl.net/#1}{\nolinkurl{http://ascl.net/#1}}}
\providecommand{\doarXiv}[1]{\href{https://arxiv.org/abs/#1}{\nolinkurl{https://arxiv.org/abs/#1}}}

\bibitem[{{Allen} {et~al.}(2011){Allen}, {Evrard}, \& {Mantz}}]{allen11}
{Allen}, S.~W., {Evrard}, A.~E., \& {Mantz}, A.~B. 2011, \araa, 49, 409,
  \dodoi{10.1146/annurev-astro-081710-102514}

\bibitem[{{Baxter} {et~al.}(2015){Baxter}, {Keisler}, {Dodelson}, {Aird},
  {Allen}, {Ashby}, {Bautz}, {Bayliss}, {Benson}, {Bleem}, {Bocquet},
  {Brodwin}, {Carlstrom}, {Chang}, {Chiu}, {Cho}, {Clocchiatti}, {Crawford},
  {Crites}, {Desai}, {Dietrich}, {de Haan}, {Dobbs}, {Foley}, {Forman},
  {George}, {Gladders}, {Gonzalez}, {Halverson}, {Harrington}, {Hennig},
  {Hoekstra}, {Holder}, {Holzapfel}, {Hou}, {Hrubes}, {Jones}, {Knox}, {Lee},
  {Leitch}, {Liu}, {Lueker}, {Luong-Van}, {Mantz}, {Marrone}, {McDonald},
  {McMahon}, {Meyer}, {Millea}, {Mocanu}, {Murray}, {Padin}, {Pryke},
  {Reichardt}, {Rest}, {Ruhl}, {Saliwanchik}, {Saro}, {Sayre}, {Schaffer},
  {Shirokoff}, {Song}, {Spieler}, {Stalder}, {Stanford}, {Staniszewski},
  {Stark}, {Story}, {van Engelen}, {Vanderlinde}, {Vieira}, {Vikhlinin},
  {Williamson}, {Zahn}, \& {Zenteno}}]{baxter15}
{Baxter}, E.~J., {Keisler}, R., {Dodelson}, S., {et~al.} 2015, \apj, 806, 247,
  \dodoi{10.1088/0004-637X/806/2/247}

\bibitem[{{Baxter} {et~al.}(2018){Baxter}, {Raghunathan}, {Crawford},
  {Fosalba}, {Hou}, {Holder}, {Omori}, {Patil}, {Rozo}, {Abbott}, {Annis},
  {Aylor}, {Benoit-L{\'e}vy}, {Benson}, {Bertin}, {Bleem}, {Buckley-Geer},
  {Burke}, {Carlstrom}, {Carnero Rosell}, {Carrasco Kind}, {Carretero},
  {Chang}, {Cho}, {Crites}, {Crocce}, {Cunha}, {da Costa}, {D'Andrea}, {Davis},
  {de Haan}, {Desai}, {Dietrich}, {Dobbs}, {Dodelson}, {Doel}, {Drlica-Wagner},
  {Estrada}, {Everett}, {Fausti Neto}, {Flaugher}, {Frieman},
  {Garc{\'{\i}}a-Bellido}, {George}, {Gaztanaga}, {Giannantonio}, {Gruen},
  {Gruendl}, {Gschwend}, {Gutierrez}, {Halverson}, {Harrington}, {Hartley},
  {Holzapfel}, {Honscheid}, {Hrubes}, {Jain}, {James}, {Jarvis}, {Jeltema},
  {Knox}, {Krause}, {Kuehn}, {Kuhlmann}, {Kuropatkin}, {Lahav}, {Lee},
  {Leitch}, {Li}, {Lima}, {Luong-Van}, {Manzotti}, {March}, {Marrone},
  {Marshall}, {Martini}, {McMahon}, {Melchior}, {Menanteau}, {Meyer}, {Miller},
  {Miquel}, {Mocanu}, {Mohr}, {Natoli}, {Nord}, {Ogando}, {Padin}, {Plazas},
  {Pryke}, {Rapetti}, {Reichardt}, {Romer}, {Roodman}, {Ruhl}, {Rykoff},
  {Sako}, {Sanchez}, {Sayre}, {Scarpine}, {Schaffer}, {Schindler}, {Schubnell},
  {Sevilla-Noarbe}, {Shirokoff}, {Smith}, {Smith}, {Soares-Santos}, {Sobreira},
  {Staniszewski}, {Stark}, {Story}, {Suchyta}, {Tarle}, {Thomas}, {Troxel},
  {Vanderlinde}, {Vieira}, {Walker}, {Williamson}, {Zhang}, \&
  {Zuntz}}]{baxter18}
{Baxter}, E.~J., {Raghunathan}, S., {Crawford}, T.~M., {et~al.} 2018, \mnras,
  476, 2674, \dodoi{10.1093/mnras/sty305}

\bibitem[{{Benson} {et~al.}(2014){Benson}, {Ade}, {Ahmed}, {Allen}, {Arnold},
  {Austermann}, {Bender}, {Bleem}, {Carlstrom}, {Chang}, {Cho}, {Ciocys},
  {Cliche}, {Crawford}, {Cukierman}, {de Haan}, {Dobbs}, {Dutcher}, {Everett},
  {Gilbert}, {Halverson}, {Hanson}, {Harrington}, {Hattori}, {Henning},
  {Hilton}, {Holder}, {Holzapfel}, {Irwin}, {Keisler}, {Knox}, {Kubik}, {Kuo},
  {Lee}, {Leitch}, {Li}, {McDonald}, {Meyer}, {Montgomery}, {Myers}, {Natoli},
  {Nguyen}, {Novosad}, {Padin}, {Pan}, {Pearson}, {Reichardt}, {Ruhl},
  {Saliwanchik}, {Simard}, {Smecher}, {Sayre}, {Shirokoff}, {Stark}, {Story},
  {Suzuki}, {Thompson}, {Tucker}, {Vanderlinde}, {Vieira}, {Vikhlinin}, {Wang},
  {Yefremenko}, \& {Yoon}}]{benson14}
{Benson}, B.~A., {Ade}, P.~A.~R., {Ahmed}, Z., {et~al.} 2014, in Society of
  Photo-Optical Instrumentation Engineers (SPIE) Conference Series, Vol. 9153,
  Society of Photo-Optical Instrumentation Engineers (SPIE) Conference Series.
\newblock \doarXiv{1407.2973}

\bibitem[{{Bleem} {et~al.}(2015){Bleem}, {Stalder}, {de Haan}, {Aird}, {Allen},
  {Applegate}, {Ashby}, {Bautz}, {Bayliss}, {Benson}, {Bocquet}, {Brodwin},
  {Carlstrom}, {Chang}, {Chiu}, {Cho}, {Clocchiatti}, {Crawford}, {Crites},
  {Desai}, {Dietrich}, {Dobbs}, {Foley}, {Forman}, {George}, {Gladders},
  {Gonzalez}, {Halverson}, {Hennig}, {Hoekstra}, {Holder}, {Holzapfel},
  {Hrubes}, {Jones}, {Keisler}, {Knox}, {Lee}, {Leitch}, {Liu}, {Lueker},
  {Luong-Van}, {Mantz}, {Marrone}, {McDonald}, {McMahon}, {Meyer}, {Mocanu},
  {Mohr}, {Murray}, {Padin}, {Pryke}, {Reichardt}, {Rest}, {Ruel}, {Ruhl},
  {Saliwanchik}, {Saro}, {Sayre}, {Schaffer}, {Schrabback}, {Shirokoff},
  {Song}, {Spieler}, {Stanford}, {Staniszewski}, {Stark}, {Story}, {Stubbs},
  {Vanderlinde}, {Vieira}, {Vikhlinin}, {Williamson}, {Zahn}, \&
  {Zenteno}}]{bleem15b}
{Bleem}, L.~E., {Stalder}, B., {de Haan}, T., {et~al.} 2015, \apjs, 216, 27,
  \dodoi{10.1088/0067-0049/216/2/27}

\bibitem[{{Bocquet} {et~al.}(2019){Bocquet}, {Dietrich}, {Schrabback}, {Bleem},
  {Klein}, {Allen}, {Applegate}, {Ashby}, {Bautz}, {Bayliss}, {Benson},
  {Brodwin}, {Bulbul}, {Canning}, {Capasso}, {Carlstrom}, {Chang}, {Chiu},
  {Cho}, {Clocchiatti}, {Crawford}, {Crites}, {de Haan}, {Desai}, {Dobbs},
  {Foley}, {Forman}, {Garmire}, {George}, {Gladders}, {Gonzalez}, {Grandis},
  {Gupta}, {Halverson}, {Hlavacek-Larrondo}, {Hoekstra}, {Holder}, {Holzapfel},
  {Hou}, {Hrubes}, {Huang}, {Jones}, {Khullar}, {Knox}, {Kraft}, {Lee}, {von
  der Linden}, {Luong-Van}, {Mantz}, {Marrone}, {McDonald}, {McMahon}, {Meyer},
  {Mocanu}, {Mohr}, {Morris}, {Padin}, {Patil}, {Pryke}, {Rapetti},
  {Reichardt}, {Rest}, {Ruhl}, {Saliwanchik}, {Saro}, {Sayre}, {Schaffer},
  {Shirokoff}, {Stalder}, {Stanford}, {Staniszewski}, {Stark}, {Story},
  {Strazzullo}, {Stubbs}, {Vanderlinde}, {Vieira}, {Vikhlinin}, {Williamson},
  \& {Zenteno}}]{bocquet19}
{Bocquet}, S., {Dietrich}, J.~P., {Schrabback}, T., {et~al.} 2019, \apj, 878,
  55, \dodoi{10.3847/1538-4357/ab1f10}

\bibitem[{{Cash}(1979)}]{cash79}
{Cash}, W. 1979, \apj, 228, 939, \dodoi{10.1086/156922}

\bibitem[{{CMB-S4 Collaboration}(2019)}]{cmbs4collab19}
{CMB-S4 Collaboration}. 2019, arXiv e-prints, arXiv:1907.04473.
\newblock \doarXiv{1907.04473}

\bibitem[{{de Haan} {et~al.}(2016){de Haan}, {Benson}, {Bleem}, {Allen},
  {Applegate}, {Ashby}, {Bautz}, {Bayliss}, {Bocquet}, {Brodwin}, {Carlstrom},
  {Chang}, {Chiu}, {Cho}, {Clocchiatti}, {Crawford}, {Crites}, {Desai},
  {Dietrich}, {Dobbs}, {Doucouliagos}, {Foley}, {Forman}, {Garmire}, {George},
  {Gladders}, {Gonzalez}, {Gupta}, {Halverson}, {Hlavacek-Larrondo},
  {Hoekstra}, {Holder}, {Holzapfel}, {Hou}, {Hrubes}, {Huang}, {Jones},
  {Keisler}, {Knox}, {Lee}, {Leitch}, {von der Linden}, {Luong-Van}, {Mantz},
  {Marrone}, {McDonald}, {McMahon}, {Meyer}, {Mocanu}, {Mohr}, {Murray},
  {Padin}, {Pryke}, {Rapetti}, {Reichardt}, {Rest}, {Ruel}, {Ruhl},
  {Saliwanchik}, {Saro}, {Sayre}, {Schaffer}, {Schrabback}, {Shirokoff},
  {Song}, {Spieler}, {Stalder}, {Stanford}, {Staniszewski}, {Stark}, {Story},
  {Stubbs}, {Vanderlinde}, {Vieira}, {Vikhlinin}, {Williamson}, \&
  {Zenteno}}]{dehaan16}
{de Haan}, T., {Benson}, B.~A., {Bleem}, L.~E., {et~al.} 2016, \apj, 832, 95,
  \dodoi{10.3847/0004-637X/832/1/95}

\bibitem[{{Dietrich} {et~al.}(2019){Dietrich}, {Bocquet}, {Schrabback},
  {Applegate}, {Hoekstra}, {Grandis}, {Mohr}, {Allen}, {Bayliss}, {Benson},
  {Bleem}, {Brodwin}, {Bulbul}, {Capasso}, {Chiu}, {Crawford}, {Gonzalez}, {de
  Haan}, {Klein}, {von der Linden}, {Mantz}, {Marrone}, {McDonald},
  {Raghunathan}, {Rapetti}, {Reichardt}, {Saro}, {Stalder}, {Stark}, {Stern},
  \& {Stubbs}}]{dietrich19}
{Dietrich}, J.~P., {Bocquet}, S., {Schrabback}, T., {et~al.} 2019, \mnras, 483,
  2871, \dodoi{10.1093/mnras/sty3088}

\bibitem[{{Dodelson}(2004)}]{dodelson04}
{Dodelson}, S. 2004, \prd, 70, 023009, \dodoi{10.1103/PhysRevD.70.023009}

\bibitem[{{Euclid Collaboration} {et~al.}(2019){Euclid Collaboration}, {Adam},
  {Vannier}, {Maurogordato}, {Biviano}, {Adami}, {Ascaso}, {Bellagamba},
  {Benoist}, {Cappi}, {D{\'\i}az-S{\'a}nchez}, {Durret}, {Farrens}, {Gonzalez},
  {Iovino}, {Licitra}, {Maturi}, {Mei}, {Merson}, {Munari}, {Pell{\'o}},
  {Ricci}, {Rocci}, {Roncarelli}, {Sarron}, {Amoura}, {Andreon}, {Apostolakos},
  {Arnaud}, {Bardelli}, {Bartlett}, {Baugh}, {Borgani}, {Brodwin}, {Castander},
  {Castignani}, {Cucciati}, {De Lucia}, {Dubath}, {Fosalba}, {Giocoli},
  {Hoekstra}, {Mamon}, {Melin}, {Moscardini}, {Paltani}, {Radovich},
  {Sartoris}, {Schultheis}, {Sereno}, {Weller}, {Burigana}, {Carvalho},
  {Corcione}, {Kurki-Suonio}, {Lilje}, {Sirri}, {Toledo-Moreo}, \&
  {Zamorani}}]{euclid19}
{Euclid Collaboration}, {Adam}, R., {Vannier}, M., {et~al.} 2019, Astronomy and
  Astrophysics, 627, A23, \dodoi{10.1051/0004-6361/201935088}

\bibitem[{Feroz {et~al.}(2009)Feroz, Hobson, \& Bridges}]{feroz08}
Feroz, F., Hobson, M., \& Bridges, M. 2009, Mon. Not. Roy. Astron. Soc., 398,
  1601, \dodoi{10.1111/j.1365-2966.2009.14548.x}

\bibitem[{{Foreman-Mackey} {et~al.}(2013){Foreman-Mackey}, {Hogg}, {Lang}, \&
  {Goodman}}]{foreman13}
{Foreman-Mackey}, D., {Hogg}, D.~W., {Lang}, D., \& {Goodman}, J. 2013, \pasp,
  125, 306, \dodoi{10.1086/670067}

\bibitem[{Grandis {et~al.}(2019)Grandis, Mohr, Dietrich, Bocquet, Saro, Klein,
  Paulus, \& Capasso}]{grandis18}
Grandis, S., Mohr, J.~J., Dietrich, J.~P., {et~al.} 2019, Monthly Notices of
  the Royal Astronomical Society, 488, 2041, \dodoi{10.1093/mnras/stz1778}

\bibitem[{{Gupta} \& {Reichardt}(2020)}]{gupta20b}
{Gupta}, N., \& {Reichardt}, C.~L. 2020, arXiv e-prints, arXiv:2005.13985.
\newblock \doarXiv{2005.13985}

\bibitem[{{Haiman} {et~al.}(2001){Haiman}, {Mohr}, \& {Holder}}]{haiman01}
{Haiman}, Z., {Mohr}, J.~J., \& {Holder}, G.~P. 2001, \apj, 553, 545,
  \dodoi{10.1086/320939}

\bibitem[{{Holder}(2006)}]{holder06}
{Holder}, G. 2006, arXiv e-prints, astro.
\newblock \doarXiv{astro-ph/0602251}

\bibitem[{{Holder} \& {Kosowsky}(2004)}]{holder04}
{Holder}, G., \& {Kosowsky}, A. 2004, \apj, 616, 8, \dodoi{10.1086/424808}

\bibitem[{{Hu} {et~al.}(2007){Hu}, {DeDeo}, \& {Vale}}]{hu07}
{Hu}, W., {DeDeo}, S., \& {Vale}, C. 2007, New Journal of Physics, 9, 441,
  \dodoi{10.1088/1367-2630/9/12/441}

\bibitem[{Hunter(2007)}]{hunter07}
Hunter, J.~D. 2007, Computing In Science \& Engineering, 9, 90,
  \dodoi{10.1109/MCSE.2007.55}

\bibitem[{Jones {et~al.}(2001)Jones, Oliphant, Peterson, {et~al.}}]{jones01}
Jones, E., Oliphant, T., Peterson, P., {et~al.} 2001, {SciPy}: Open source
  scientific tools for {Python}.
\newblock \url{http://www.scipy.org/}

\bibitem[{{Lafayette} {et~al.}(2016){Lafayette}, {Sauter}, {Vu}, \&
  {Meade}}]{spartan}
{Lafayette}, L., {Sauter}, G., {Vu}, L., \& {Meade}, B. 2016, OpenStack Summit,
  Barcelona, \dodoi{10.4225/49/58ead90dceaaa}

\bibitem[{{Lewis} \& {Challinor}(2006)}]{lewis06}
{Lewis}, A., \& {Challinor}, A. 2006, \physrep, 429, 1,
  \dodoi{10.1016/j.physrep.2006.03.002}

\bibitem[{Lewis {et~al.}(2000)Lewis, Challinor, \& Lasenby}]{lewis99}
Lewis, A., Challinor, A., \& Lasenby, A. 2000, Astrophys. J., 538, 473

\bibitem[{{Lewis} \& {King}(2006)}]{lewis06a}
{Lewis}, A., \& {King}, L. 2006, \prd, 73, 063006,
  \dodoi{10.1103/PhysRevD.73.063006}

\bibitem[{{LSST Science Collaboration} {et~al.}(2009){LSST Science
  Collaboration}, {Abell}, {Allison}, {Anderson}, {Andrew}, {Angel}, {Armus},
  {Arnett}, {Asztalos}, {Axelrod}, {Bailey}, {Ballantyne}, {Bankert},
  {Barkhouse}, {Barr}, {Barrientos}, {Barth}, {Bartlett}, {Becker}, {Becla},
  {Beers}, {Bernstein}, {Biswas}, {Blanton}, {Bloom}, {Bochanski}, {Boeshaar},
  {Borne}, {Bradac}, {Brandt}, {Bridge}, {Brown}, {Brunner}, {Bullock},
  {Burgasser}, {Burge}, {Burke}, {Cargile}, {Chand rasekharan}, {Chartas},
  {Chesley}, {Chu}, {Cinabro}, {Claire}, {Claver}, {Clowe}, {Connolly}, {Cook},
  {Cooke}, {Cooray}, {Covey}, {Culliton}, {de Jong}, {de Vries}, {Debattista},
  {Delgado}, {Dell'Antonio}, {Dhital}, {Di Stefano}, {Dickinson}, {Dilday},
  {Djorgovski}, {Dobler}, {Donalek}, {Dubois-Felsmann}, {Durech},
  {Eliasdottir}, {Eracleous}, {Eyer}, {Falco}, {Fan}, {Fassnacht}, {Ferguson},
  {Fernandez}, {Fields}, {Finkbeiner}, {Figueroa}, {Fox}, {Francke}, {Frank},
  {Frieman}, {Fromenteau}, {Furqan}, {Galaz}, {Gal-Yam}, {Garnavich},
  {Gawiser}, {Geary}, {Gee}, {Gibson}, {Gilmore}, {Grace}, {Green}, {Gressler},
  {Grillmair}, {Habib}, {Haggerty}, {Hamuy}, {Harris}, {Hawley}, {Heavens},
  {Hebb}, {Henry}, {Hileman}, {Hilton}, {Hoadley}, {Holberg}, {Holman},
  {Howell}, {Infante}, {Ivezic}, {Jacoby}, {Jain}, {R}, {Jedicke}, {Jee},
  {Garrett Jernigan}, {Jha}, {Johnston}, {Jones}, {Juric}, {Kaasalainen},
  {Styliani}, {Kafka}, {Kahn}, {Kaib}, {Kalirai}, {Kantor}, {Kasliwal},
  {Keeton}, {Kessler}, {Knezevic}, {Kowalski}, {Krabbendam}, {Krughoff},
  {Kulkarni}, {Kuhlman}, {Lacy}, {Lepine}, {Liang}, {Lien}, {Lira}, {Long},
  {Lorenz}, {Lotz}, {Lupton}, {Lutz}, {Macri}, {Mahabal}, {Mandelbaum},
  {Marshall}, {May}, {McGehee}, {Meadows}, {Meert}, {Milani}, {Miller},
  {Miller}, {Mills}, {Minniti}, {Monet}, {Mukadam}, {Nakar}, {Neill}, {Newman},
  {Nikolaev}, {Nordby}, {O'Connor}, {Oguri}, {Oliver}, {Olivier}, {Olsen},
  {Olsen}, {Olszewski}, {Oluseyi}, {Padilla}, {Parker}, {Pepper}, {Peterson},
  {Petry}, {Pinto}, {Pizagno}, {Popescu}, {Prsa}, {Radcka}, {Raddick},
  {Rasmussen}, {Rau}, {Rho}, {Rhoads}, {Richards}, {Ridgway}, {Robertson},
  {Roskar}, {Saha}, {Sarajedini}, {Scannapieco}, {Schalk}, {Schindler},
  {Schmidt}, {Schmidt}, {Schneider}, {Schumacher}, {Scranton}, {Sebag},
  {Seppala}, {Shemmer}, {Simon}, {Sivertz}, {Smith}, {Allyn Smith}, {Smith},
  {Spitz}, {Stanford}, {Stassun}, {Strader}, {Strauss}, {Stubbs}, {Sweeney},
  {Szalay}, {Szkody}, {Takada}, {Thorman}, {Trilling}, {Trimble}, {Tyson}, {Van
  Berg}, {Vand en Berk}, {VanderPlas}, {Verde}, {Vrsnak}, {Walkowicz}, {Wand
  elt}, {Wang}, {Wang}, {Warner}, {Wechsler}, {West}, {Wiecha}, {Williams},
  {Willman}, {Wittman}, {Wolff}, {Wood-Vasey}, {Wozniak}, {Young}, {Zentner},
  \& {Zhan}}]{lsst09}
{LSST Science Collaboration}, {Abell}, P.~A., {Allison}, J., {et~al.} 2009,
  arXiv e-prints, arXiv:0912.0201.
\newblock \doarXiv{0912.0201}

\bibitem[{{Madhavacheril} {et~al.}(2015){Madhavacheril}, {Sehgal}, {Allison},
  {Battaglia}, {Bond}, {Calabrese}, {Caliguiri}, {Coughlin}, {Crichton},
  {Datta}, {Devlin}, {Dunkley}, {D{\"u}nner}, {Fogarty}, {Grace}, {Hajian},
  {Hasselfield}, {Hill}, {Hilton}, {Hincks}, {Hlozek}, {Hughes}, {Kosowsky},
  {Louis}, {Lungu}, {McMahon}, {Moodley}, {Munson}, {Naess}, {Nati},
  {Newburgh}, {Niemack}, {Page}, {Partridge}, {Schmitt}, {Sherwin}, {Sievers},
  {Spergel}, {Staggs}, {Thornton}, {Van Engelen}, {Ward}, {Wollack}, \&
  {Atacama Cosmology Telescope Collaboration}}]{madhavacheril15}
{Madhavacheril}, M., {Sehgal}, N., {Allison}, R., {et~al.} 2015, Physical
  Review Letters, 114, 151302, \dodoi{10.1103/PhysRevLett.114.151302}

\bibitem[{{McClintock} {et~al.}(2019){McClintock}, {Varga}, {Gruen}, {Rozo},
  {Rykoff}, {Shin}, {Melchior}, {DeRose}, {Seitz}, {Dietrich}, {Sheldon},
  {Zhang}, {von der Linden}, {Jeltema}, {Mantz}, {Romer}, {Allen}, {Becker},
  {Bermeo}, {Bhargava}, {Costanzi}, {Everett}, {Farahi}, {Hamaus}, {Hartley},
  {Hollowood}, {Hoyle}, {Israel}, {Li}, {MacCrann}, {Morris}, {Palmese},
  {Plazas}, {Pollina}, {Rau}, {Simet}, {Soares-Santos}, {Troxel}, {Vergara
  Cervantes}, {Wechsler}, {Zuntz}, {Abbott}, {Abdalla}, {Allam}, {Annis},
  {Avila}, {Bridle}, {Brooks}, {Burke}, {Carnero Rosell}, {Carrasco Kind},
  {Carretero}, {Castander}, {Crocce}, {Cunha}, {D'Andrea}, {da Costa}, {Davis},
  {De Vicente}, {Diehl}, {Doel}, {Drlica-Wagner}, {Evrard}, {Flaugher},
  {Fosalba}, {Frieman}, {Garc{\'\i}a-Bellido}, {Gaztanaga}, {Gerdes},
  {Giannantonio}, {Gruendl}, {Gutierrez}, {Honscheid}, {James}, {Kirk},
  {Krause}, {Kuehn}, {Lahav}, {Li}, {Lima}, {March}, {Marshall}, {Menanteau},
  {Miquel}, {Mohr}, {Nord}, {Ogando}, {Roodman}, {Sanchez}, {Scarpine},
  {Schindler}, {Sevilla-Noarbe}, {Smith}, {Smith}, {Sobreira}, {Suchyta},
  {Swanson}, {Tarle}, {Tucker}, {Vikram}, {Walker}, \& {Weller}}]{mcclintock19}
{McClintock}, T., {Varga}, T.~N., {Gruen}, D., {et~al.} 2019, \mnras, 482,
  1352, \dodoi{10.1093/mnras/sty2711}

\bibitem[{{McDonald} {et~al.}(2013){McDonald}, {Benson}, {Vikhlinin},
  {Stalder}, {Bleem}, {de Haan}, {Lin}, {Aird}, {Ashby}, {Bautz}, {Bayliss},
  {Bocquet}, {Brodwin}, {Carlstrom}, {Chang}, {Cho}, {Clocchiatti}, {Crawford},
  {Crites}, {Desai}, {Dobbs}, {Dudley}, {Foley}, {Forman}, {George},
  {Gettings}, {Gladders}, {Gonzalez}, {Halverson}, {High}, {Holder},
  {Holzapfel}, {Hoover}, {Hrubes}, {Jones}, {Joy}, {Keisler}, {Knox}, {Lee},
  {Leitch}, {Liu}, {Lueker}, {Luong-Van}, {Mantz}, {Marrone}, {McMahon},
  {Mehl}, {Meyer}, {Miller}, {Mocanu}, {Mohr}, {Montroy}, {Murray},
  {Nurgaliev}, {Padin}, {Plagge}, {Pryke}, {Reichardt}, {Rest}, {Ruel}, {Ruhl},
  {Saliwanchik}, {Saro}, {Sayre}, {Schaffer}, {Shirokoff}, {Song}, {{\v
  S}uhada}, {Spieler}, {Stanford}, {Staniszewski}, {Stark}, {Story}, {van
  Engelen}, {Vanderlinde}, {Vieira}, {Williamson}, {Zahn}, \&
  {Zenteno}}]{mcdonald13}
{McDonald}, M., {Benson}, B.~A., {Vikhlinin}, A., {et~al.} 2013, \apj, 774, 23,
  \dodoi{10.1088/0004-637X/774/1/23}

\bibitem[{{McDonald} {et~al.}(2017){McDonald}, {Allen}, {Bayliss}, {Benson},
  {Bleem}, {Brodwin}, {Bulbul}, {Carlstrom}, {Forman}, {Hlavacek-Larrondo},
  {Garmire}, {Gaspari}, {Gladders}, {Mantz}, \& {Murray}}]{mcdonald17}
{McDonald}, M., {Allen}, S.~W., {Bayliss}, M., {et~al.} 2017, \apj, 843, 28,
  \dodoi{10.3847/1538-4357/aa7740}

\bibitem[{{Melin} \& {Bartlett}(2015)}]{melin15}
{Melin}, J.-B., \& {Bartlett}, J.~G. 2015, \aap, 578, A21,
  \dodoi{10.1051/0004-6361/201424720}

\bibitem[{{Merloni} {et~al.}(2012){Merloni}, {Predehl}, {Becker},
  {B{\"o}hringer}, {Boller}, {Brunner}, {Brusa}, {Dennerl}, {Freyberg}, \&
  {Friedrich}}]{merloni12}
{Merloni}, A., {Predehl}, P., {Becker}, W., {et~al.} 2012, arXiv e-prints,
  arXiv:1209.3114.
\newblock \doarXiv{1209.3114}

\bibitem[{{Navarro} {et~al.}(1997){Navarro}, {Frenk}, \& {White}}]{navarro97}
{Navarro}, J.~F., {Frenk}, C.~S., \& {White}, S.~D.~M. 1997, \apj, 490, 493,
  \dodoi{10.1086/304888}

\bibitem[{{Planck Collaboration} {et~al.}(2016){Planck Collaboration}, {Ade},
  {Aghanim}, {Arnaud}, {Ashdown}, {Aumont}, {Baccigalupi}, {Banday},
  {Barreiro}, {Bartlett}, \& et~al.}]{planck15-24}
{Planck Collaboration}, {Ade}, P.~A.~R., {Aghanim}, N., {et~al.} 2016, \aap,
  594, A24, \dodoi{10.1051/0004-6361/201525833}

\bibitem[{{Planck Collaboration} {et~al.}(2020){Planck Collaboration},
  {Aghanim}, {Akrami}, {Ashdown}, {Aumont}, {Baccigalupi}, {Ballardini},
  {Banday}, {Barreiro}, {Bartolo}, {Basak}, {Benabed}, {Bernard}, {Bersanelli},
  {Bielewicz}, {Bock}, {Bond}, {Borrill}, {Bouchet}, {Boulanger}, {Bucher},
  {Burigana}, {Butler}, {Calabrese}, {Cardoso}, {Carron}, {Casaponsa},
  {Challinor}, {Chiang}, {Colombo}, {Combet}, {Crill}, {Cuttaia}, {de
  Bernardis}, {de Rosa}, {de Zotti}, {Delabrouille}, {Delouis}, {Di Valentino},
  {Diego}, {Dor{\'e}}, {Douspis}, {Ducout}, {Dupac}, {Dusini}, {Efstathiou},
  {Elsner}, {En{\ss}lin}, {Eriksen}, {Fantaye}, {Fernand ez-Cobos}, {Finelli},
  {Frailis}, {Fraisse}, {Franceschi}, {Frolov}, {Galeotta}, {Galli}, {Ganga},
  {G{\'e}nova-Santos}, {Gerbino}, {Ghosh}, {Giraud-H{\'e}raud},
  {Gonz{\'a}lez-Nuevo}, {G{\'o}rski}, {Gratton}, {Gruppuso}, {Gudmundsson},
  {Hamann}, {Handley}, {Hansen}, {Herranz}, {Hivon}, {Huang}, {Jaffe}, {Jones},
  {Keih{\"a}nen}, {Keskitalo}, {Kiiveri}, {Kim}, {Kisner}, {Krachmalnicoff},
  {Kunz}, {Kurki-Suonio}, {Lagache}, {Lamarre}, {Lasenby}, {Lattanzi},
  {Lawrence}, {Le Jeune}, {Levrier}, {Lewis}, {Liguori}, {Lilje}, {Lilley},
  {Lindholm}, {L{\'o}pez-Caniego}, {Lubin}, {Ma}, {Mac{\'\i}as-P{\'e}rez},
  {Maggio}, {Maino}, {Mandolesi}, {Mangilli}, {Marcos-Caballero}, {Maris},
  {Martin}, {Mart{\'\i}nez-Gonz{\'a}lez}, {Matarrese}, {Mauri}, {McEwen},
  {Meinhold}, {Melchiorri}, {Mennella}, {Migliaccio}, {Millea},
  {Miville-Desch{\^e}nes}, {Molinari}, {Moneti}, {Montier}, {Morgante}, {Moss},
  {Natoli}, {N{\o}rgaard-Nielsen}, {Pagano}, {Paoletti}, {Partridge},
  {Patanchon}, {Peiris}, {Perrotta}, {Pettorino}, {Piacentini}, {Polenta},
  {Puget}, {Rachen}, {Reinecke}, {Remazeilles}, {Renzi}, {Rocha}, {Rosset},
  {Roudier}, {Rubi{\~n}o-Mart{\'\i}n}, {Ruiz-Granados}, {Salvati}, {Sandri},
  {Savelainen}, {Scott}, {Shellard}, {Sirignano}, {Sirri}, {Spencer},
  {Sunyaev}, {Suur-Uski}, {Tauber}, {Tavagnacco}, {Tenti}, {Toffolatti},
  {Tomasi}, {Trombetti}, {Valiviita}, {Van Tent}, {Vielva}, {Villa},
  {Vittorio}, {Wandelt}, {Wehus}, {Zacchei}, \& {Zonca}}]{planck18-5}
{Planck Collaboration}, {Aghanim}, N., {Akrami}, Y., {et~al.} 2020, \aap, 641,
  A5, \dodoi{10.1051/0004-6361/201936386}

\bibitem[{{Raghunathan} {et~al.}(2017){Raghunathan}, {Patil}, {Baxter},
  {Bianchini}, {Bleem}, {Crawford}, {Holder}, {Manzotti}, \&
  {Reichardt}}]{raghunathan17}
{Raghunathan}, S., {Patil}, S., {Baxter}, E.~J., {et~al.} 2017, \jcap, 8, 030,
  \dodoi{10.1088/1475-7516/2017/08/030}

\bibitem[{{Raghunathan} {et~al.}(2019{\natexlab{a}}){Raghunathan}, {Patil},
  {Baxter}, {Benson}, {Bleem}, {Crawford}, {Holder}, {McClintock}, {Reichardt},
  {Varga}, {Whitehorn}, {Ade}, {Allam}, {Anderson}, {Austermann}, {Avila},
  {Avva}, {Bacon}, {Beall}, {Bender}, {Bianchini}, {Bocquet}, {Brooks},
  {Burke}, {Carlstrom}, {Carretero}, {Castander}, {Chang}, {Chiang}, {Citron},
  {Costanzi}, {Crites}, {da Costa}, {Desai}, {Diehl}, {Dietrich}, {Dobbs},
  {Doel}, {Everett}, {Evrard}, {Feng}, {Flaugher}, {Fosalba}, {Frieman},
  {Gallicchio}, {Garc{\'\i}a-Bellido}, {Gaztanaga}, {George}, {Giannantonio},
  {Gilbert}, {Gruendl}, {Gschwend}, {Gupta}, {Gutierrez}, {de Haan},
  {Halverson}, {Harrington}, {Henning}, {Hilton}, {Hollowood}, {Holzapfel},
  {Honscheid}, {Hrubes}, {Huang}, {Hubmayr}, {Irwin}, {Jeltema}, {Kind},
  {Knox}, {Kuropatkin}, {Lahav}, {Lee}, {Li}, {Lima}, {Lowitz}, {Maia},
  {Marshall}, {McMahon}, {Melchior}, {Menanteau}, {Meyer}, {Miquel}, {Mocanu},
  {Mohr}, {Montgomery}, {Moran}, {Nadolski}, {Natoli}, {Nibarger}, {Noble},
  {Novosad}, {Ogand o}, {Padin}, {Plazas}, {Pryke}, {Rapetti}, {Romer},
  {Roodman}, {Rosell}, {Rozo}, {Ruhl}, {Rykoff}, {Saliwanchik}, {Sanchez},
  {Sayre}, {Scarpine}, {Schaffer}, {Schubnell}, {Serrano}, {Sevilla-Noarbe},
  {Sievers}, {Smecher}, {Smith}, {Soares-Santos}, {Stark}, {Story}, {Suchyta},
  {Swanson}, {Tarle}, {Tucker}, {Vanderlinde}, {Veach}, {De Vicente}, {Vieira},
  {Vikram}, {Wang}, {Wu}, {Yefremenko}, {Zhang}, {SPTpol}, \& {DES
  Collaboration}}]{raghunathan19b}
{Raghunathan}, S., {Patil}, S., {Baxter}, E., {et~al.} 2019{\natexlab{a}},
  \prl, 123, 181301, \dodoi{10.1103/PhysRevLett.123.181301}

\bibitem[{{Raghunathan} {et~al.}(2019{\natexlab{b}}){Raghunathan}, {Patil},
  {Baxter}, {Benson}, {Bleem}, {Chou}, {Crawford}, {Holder}, {McClintock},
  {Reichardt}, {Rozo}, {Varga}, {Abbott}, {Ade}, {Allam}, {Anderson}, {Annis},
  {Austermann}, {Avila}, {Beall}, {Bechtol}, {Bender}, {Bernstein}, {Bertin},
  {Bianchini}, {Brooks}, {Burke}, {Carlstrom}, {Carretero}, {Chang}, {Chiang},
  {Cho}, {Citron}, {Crites}, {Cunha}, {da Costa}, {Davis}, {Desai}, {Diehl},
  {Dietrich}, {Dobbs}, {Doel}, {Eifler}, {Everett}, {Evrard}, {Flaugher},
  {Fosalba}, {Frieman}, {Gallicchio}, {Garc{\'{\i}}a-Bellido}, {Gaztanaga},
  {George}, {Gilbert}, {Gruen}, {Gruendl}, {Gschwend}, {Gupta}, {Gutierrez},
  {de Haan}, {Halverson}, {Harrington}, {Hartley}, {Henning}, {Hilton},
  {Hollowood}, {Holzapfel}, {Honscheid}, {Hou}, {Hoyle}, {Hrubes}, {Huang},
  {Hubmayr}, {Irwin}, {James}, {Jeltema}, {Kim}, {Carrasco Kind}, {Knox},
  {Kovacs}, {Kuehn}, {Kuropatkin}, {Lee}, {Li}, {Lima}, {Maia}, {Marshall},
  {McMahon}, {Melchior}, {Menanteau}, {Meyer}, {Miller}, {Miquel}, {Mocanu},
  {Montgomery}, {Nadolski}, {Natoli}, {Nibarger}, {Novosad}, {Padin}, {Plazas},
  {Pryke}, {Rapetti}, {Romer}, {Carnero Rosell}, {Ruhl}, {Saliwanchik},
  {Sanchez}, {Sayre}, {Scarpine}, {Schaffer}, {Schubnell}, {Serrano},
  {Sevilla-Noarbe}, {Smecher}, {Smith}, {Soares-Santos}, {Sobreira}, {Stark},
  {Story}, {Suchyta}, {Swanson}, {Tarle}, {Thomas}, {Tucker}, {Vanderlinde},
  {De Vicente}, {Vieira}, {Wang}, {Whitehorn}, {Wu}, \&
  {Zhang}}]{raghunathan19a}
---. 2019{\natexlab{b}}, \apj, 872, 170, \dodoi{10.3847/1538-4357/ab01ca}

\bibitem[{{Reichardt} {et~al.}(2013){Reichardt}, {Stalder}, {Bleem}, {Montroy},
  {Aird}, {Andersson}, {Armstrong}, {Ashby}, {Bautz}, {Bayliss}, {Bazin},
  {Benson}, {Brodwin}, {Carlstrom}, {Chang}, {Cho}, {Clocchiatti}, {Crawford},
  {Crites}, {de Haan}, {Desai}, {Dobbs}, {Dudley}, {Foley}, {Forman}, {George},
  {Gladders}, {Gonzalez}, {Halverson}, {Harrington}, {High}, {Holder},
  {Holzapfel}, {Hoover}, {Hrubes}, {Jones}, {Joy}, {Keisler}, {Knox}, {Lee},
  {Leitch}, {Liu}, {Lueker}, {Luong-Van}, {Mantz}, {Marrone}, {McDonald},
  {McMahon}, {Mehl}, {Meyer}, {Mocanu}, {Mohr}, {Murray}, {Natoli}, {Padin},
  {Plagge}, {Pryke}, {Rest}, {Ruel}, {Ruhl}, {Saliwanchik}, {Saro}, {Sayre},
  {Schaffer}, {Shaw}, {Shirokoff}, {Song}, {Spieler}, {Staniszewski}, {Stark},
  {Story}, {Stubbs}, {{\v S}uhada}, {van Engelen}, {Vanderlinde}, {Vieira},
  {Vikhlinin}, {Williamson}, {Zahn}, \& {Zenteno}}]{reichardt13}
{Reichardt}, C.~L., {Stalder}, B., {Bleem}, L.~E., {et~al.} 2013, \apj, 763,
  127, \dodoi{10.1088/0004-637X/763/2/127}

\bibitem[{{Salvati} {et~al.}(2018){Salvati}, {Douspis}, \&
  {Aghanim}}]{salvati18}
{Salvati}, L., {Douspis}, M., \& {Aghanim}, N. 2018, \aap, 614, A13,
  \dodoi{10.1051/0004-6361/201731990}

\bibitem[{{Salvati} {et~al.}(2019){Salvati}, {Douspis}, {Ritz}, {Aghanim}, \&
  {Babul}}]{salvati19}
{Salvati}, L., {Douspis}, M., {Ritz}, A., {Aghanim}, N., \& {Babul}, A. 2019,
  A\&A, 626, A27, \dodoi{10.1051/0004-6361/201935041}

\bibitem[{{Schrabback} {et~al.}(2018){Schrabback}, {Applegate}, {Dietrich},
  {Hoekstra}, {Bocquet}, {Gonzalez}, {von der Linden}, {McDonald}, {Morrison},
  {Raihan}, {Allen}, {Bayliss}, {Benson}, {Bleem}, {Chiu}, {Desai}, {Foley},
  {de Haan}, {High}, {Hilbert}, {Mantz}, {Massey}, {Mohr}, {Reichardt}, {Saro},
  {Simon}, {Stern}, {Stubbs}, \& {Zenteno}}]{schrabback18}
{Schrabback}, T., {Applegate}, D., {Dietrich}, J.~P., {et~al.} 2018, \mnras,
  474, 2635, \dodoi{10.1093/mnras/stx2666}

\bibitem[{{Seljak} \& {Zaldarriaga}(2000)}]{seljak00b}
{Seljak}, U., \& {Zaldarriaga}, M. 2000, \apj, 538, 57, \dodoi{10.1086/309098}

\bibitem[{{Shimon} {et~al.}(2011){Shimon}, {Sadeh}, \& {Rephaeli}}]{shimon11}
{Shimon}, M., {Sadeh}, S., \& {Rephaeli}, Y. 2011, \mnras, 412, 1895,
  \dodoi{10.1111/j.1365-2966.2010.18026.x}

\bibitem[{{Sobrin} {et~al.}(2021){Sobrin}, {Anderson}, {Bender}, {Benson},
  {Dutcher}, {Foster}, {Goeckner-Wald}, {Montgomery}, {Nadolski}, {Rahlin},
  {Ade}, {Ahmed}, {Anderes}, {Archipley}, {Austermann}, {Avva}, {Aylor},
  {Balkenhol}, {Barry}, {Basu Thakur}, {Benabed}, {Bianchini}, {Bleem},
  {Bouchet}, {Bryant}, {Byrum}, {Carlstrom}, {Carter}, {Cecil}, {Chang},
  {Chaubal}, {Chen}, {Cho}, {Chou}, {Cliche}, {Crawford}, {Cukierman}, {Daley},
  {de Haan}, {Denison}, {Dibert}, {Ding}, {Dobbs}, {Everett}, {Feng},
  {Ferguson}, {Fu}, {Galli}, {Gambrel}, {Gardner}, {Gualtieri}, {Guns},
  {Gupta}, {Guyser}, {Halverson}, {Harke-Hosemann}, {Harrington}, {Henning},
  {Hilton}, {Hivon}, {Holder}, {Holzapfel}, {Hood}, {Howe}, {Huang}, {Irwin},
  {Jeong}, {Jonas}, {Jones}, {Khaire}, {Knox}, {Kofman}, {Korman}, {Kubik},
  {Kuhlmann}, {Kuo}, {Lee}, {Leitch}, {Lowitz}, {Lu}, {Meyer}, {Michalik},
  {Millea}, {Natoli}, {Nguyen}, {Noble}, {Novosad}, {Omori}, {Padin}, {Pan},
  {Paschos}, {Pearson}, {Posada}, {Prabhu}, {Quan}, {Reichardt}, {Riebel},
  {Riedel}, {Rouble}, {Ruhl}, {Saliwanchik}, {Sayre}, {Schiappucci},
  {Shirokoff}, {Smecher}, {Stark}, {Stephen}, {Story}, {Suzuki}, {Tandoi},
  {Thompson}, {Thorne}, {Tucker}, {Umilta}, {Vale}, {Vanderlinde}, {Vieira},
  {Wang}, {Whitehorn}, {Wu}, {Yefremenko}, {Yoon}, \& {Young}}]{sobrin21}
{Sobrin}, J.~A., {Anderson}, A.~J., {Bender}, A.~N., {et~al.} 2021, arXiv
  e-prints, arXiv:2106.11202.
\newblock \doarXiv{2106.11202}

\bibitem[{{The LSST Dark Energy Science Collaboration} {et~al.}(2018){The LSST
  Dark Energy Science Collaboration}, {Mandelbaum}, {Eifler}, {Hlo{\v{z}}ek},
  {Collett}, {Gawiser}, {Scolnic}, {Alonso}, {Awan}, {Biswas}, {Blazek},
  {Burchat}, {Chisari}, {Dell'Antonio}, {Digel}, {Frieman}, {Goldstein},
  {Hook}, {Ivezi{\'c}}, {Kahn}, {Kamath}, {Kirkby}, {Kitching}, {Krause},
  {Leget}, {Marshall}, {Meyers}, {Miyatake}, {Newman}, {Nichol}, {Rykoff},
  {Sanchez}, {Slosar}, {Sullivan}, \& {Troxel}}]{lsst18}
{The LSST Dark Energy Science Collaboration}, {Mandelbaum}, R., {Eifler}, T.,
  {et~al.} 2018, arXiv e-prints, arXiv:1809.01669.
\newblock \doarXiv{1809.01669}

\bibitem[{{To} {et~al.}(2020){To}, {Krause}, {Rozo}, {Wu}, {Gruen}, {Wechsler},
  {Eifler}, {Rykoff}, {Costanzi}, {Becker}, {Bernstein}, {Blazek}, {Bocquet},
  {Bridle}, {Cawthon}, {Choi}, {Crocce}, {Davis}, {DeRose}, {Drlica-Wagner},
  {Elvin-Poole}, {Fang}, {Farahi}, {Friedrich}, {Gatti}, {Gaztanaga},
  {Giannantonio}, {Hartley}, {Hoyle}, {Jarvis}, {MacCrann}, {McClintock},
  {Miranda}, {Pereira}, {Park}, {Porredon}, {Prat}, {Rau}, {Ross}, {Samuroff},
  {S{\'a}nchez}, {Sevilla-Noarbe}, {Sheldon}, {Troxel}, {Varga}, {Vielzeuf},
  {Zhang}, {Zuntz}, {Abbott}, {Aguena}, {Annis}, {Avila}, {Bertin}, {Bhargava},
  {Brooks}, {Burke}, {Carnero Rosell}, {Carrasco Kind}, {Carretero}, {Chang},
  {Conselice}, {da Costa}, {Davis}, {Desai}, {Diehl}, {Dietrich}, {Everett},
  {Evrard}, {Ferrero}, {Flaugher}, {Fosalba}, {Frieman}, {Garc{\'\i}a-Bellido},
  {Gruendl}, {Gutierrez}, {Hinton}, {Hollowood}, {Huterer}, {James}, {Jeltema},
  {Kron}, {Kuehn}, {Kuropatkin}, {Lima}, {Maia}, {Marshall}, {Menanteau},
  {Miquel}, {Morgan}, {Muir}, {Myles}, {Palmese}, {Paz-Chinch{\'o}n}, {Plazas},
  {Romer}, {Roodman}, {Sanchez}, {Santiago}, {Scarpine}, {Serrano}, {Smith},
  {Suchyta}, {Swanson}, {Tarle}, {Thomas}, {Tucker}, {Weller}, \&
  {Wester}}]{to20}
{To}, C., {Krause}, E., {Rozo}, E., {et~al.} 2020, arXiv e-prints,
  arXiv:2010.01138.
\newblock \doarXiv{2010.01138}

\bibitem[{{Vale} \& {Ostriker}(2004)}]{vale04}
{Vale}, A., \& {Ostriker}, J.~P. 2004, \mnras, 353, 189,
  \dodoi{10.1111/j.1365-2966.2004.08059.x}

\bibitem[{van~der Walt {et~al.}(2011)van~der Walt, Colbert, \&
  Varoquaux}]{vanDerWalt11}
van~der Walt, S., Colbert, S., \& Varoquaux, G. 2011, Computing in Science
  Engineering, 13, 22, \dodoi{10.1109/MCSE.2011.37}

\bibitem[{{Vanderlinde} {et~al.}(2010){Vanderlinde}, {Crawford}, {de Haan},
  {Dudley}, {Shaw}, {Ade}, {Aird}, {Benson}, {Bleem}, {Brodwin}, {Carlstrom},
  {Chang}, {Crites}, {Desai}, {Dobbs}, {Foley}, {George}, {Gladders}, {Hall},
  {Halverson}, {High}, {Holder}, {Holzapfel}, {Hrubes}, {Joy}, {Keisler},
  {Knox}, {Lee}, {Leitch}, {Loehr}, {Lueker}, {Marrone}, {McMahon}, {Mehl},
  {Meyer}, {Mohr}, {Montroy}, {Ngeow}, {Padin}, {Plagge}, {Pryke}, {Reichardt},
  {Rest}, {Ruel}, {Ruhl}, {Schaffer}, {Shirokoff}, {Song}, {Spieler},
  {Stalder}, {Staniszewski}, {Stark}, {Stubbs}, {van Engelen}, {Vieira},
  {Williamson}, {Yang}, {Zahn}, \& {Zenteno}}]{vanderlinde10}
{Vanderlinde}, K., {Crawford}, T.~M., {de Haan}, T., {et~al.} 2010, \apj, 722,
  1180, \dodoi{10.1088/0004-637X/722/2/1180}

\bibitem[{{Wang} \& {Steinhardt}(1998)}]{wang98}
{Wang}, L., \& {Steinhardt}, P.~J. 1998, \apj, 508, 483

\bibitem[{{Weller} \& {Battye}(2003)}]{weller03}
{Weller}, J., \& {Battye}, R.~A. 2003, New Astronomy Review, 47, 775,
  \dodoi{10.1016/S1387-6473(03)00137-4}

\bibitem[{{Weller} {et~al.}(2002){Weller}, {Battye}, \& {Kneissl}}]{weller02}
{Weller}, J., {Battye}, R.~A., \& {Kneissl}, R. 2002, \prl, 88, 231301,
  \dodoi{10.1103/PhysRevLett.88.231301}

\bibitem[{{Zohren} {et~al.}(2019){Zohren}, {Schrabback}, {van der Burg},
  {Arnaud}, {Melin}, {van den Busch}, {Hoekstra}, \& {Klein}}]{zohren19}
{Zohren}, H., {Schrabback}, T., {van der Burg}, R. F.~J., {et~al.} 2019,
  \mnras, 488, 2523, \dodoi{10.1093/mnras/stz1838}

\bibitem[{{Zubeldia} \& {Challinor}(2019)}]{zubeldia19}
{Zubeldia}, {\'I}., \& {Challinor}, A. 2019, \mnras, 489, 401,
  \dodoi{10.1093/mnras/stz2153}

\bibitem[{Zuntz {et~al.}(2015)Zuntz, Paterno, Jennings, Rudd, Manzotti,
  Dodelson, Bridle, Sehrish, \& Kowalkowski}]{zuntz14}
Zuntz, J., Paterno, M., Jennings, E., {et~al.} 2015, Astron. Comput., 12, 45,
  \dodoi{10.1016/j.ascom.2015.05.005}

\end{thebibliography}

\end{document}